\documentclass[10pt,journal]{IEEEtran}
\IEEEoverridecommandlockouts

\usepackage{graphicx}
\usepackage{subfigure}
\usepackage{amsmath,amsthm,amsfonts,amssymb}
\usepackage{cite}
\usepackage{bm}
\usepackage{bbm}
\usepackage{url}
\usepackage{array}
\usepackage{color}
\usepackage{multirow}
\usepackage{booktabs}
\usepackage[table,xcdraw]{xcolor}
\usepackage{enumerate}
\usepackage{enumitem}

\usepackage[english]{babel}
\usepackage{algorithm}
\usepackage[noend]{algpseudocode}

\theoremstyle{plain}

\newtheorem*{rem}{Remark}
\newtheorem{sty1}{Theorem}
\newtheorem{defi}[sty1]{Definition}

\usepackage[english]{babel}
\usepackage{algorithm}
\usepackage[noend]{algpseudocode}

\allowdisplaybreaks[4]

\begin{document}
\title{Federated Edge Learning with Misaligned Over-The-Air Computation}

\author{
Yulin~Shao,~\IEEEmembership{Member,~IEEE},
Deniz~G\"und\"uz,~\IEEEmembership{Senior Member,~IEEE},
Soung Chang Liew,~\IEEEmembership{Fellow,~IEEE}
\thanks{Y. Shao was with the Department of Information Engineering, The Chinese University of Hong Kong, Shatin, New Territories, Hong Kong. He is now with the Department of Electrical and Electronic Engineering, Imperial College London, London SW7 2AZ, U.K. (e-mail: yshao@ic.ac.uk).
D. G\"und\"uz is with the Department of Electrical and Electronic Engineering, Imperial College London, London SW7 2AZ, U.K. (e-mail: d.gunduz@imperial.ac.uk). S. C. Liew is with the Department of Information Engineering, The Chinese University of Hong Kong, Shatin, New Territories, Hong Kong (e-mail: soung@ie.cuhk.edu.hk).}
\thanks{This work was supported by the European Research Council project BEACON under grant number 677854, and by CHIST-ERA grant CHIST-ERA-18-SDCDN-001 (funded by EPSRC-EP/T023600/1).}
}

\maketitle

\begin{abstract}
Over-the-air computation (OAC) is a promising technique to realize fast model aggregation in the uplink of federated edge learning (FEEL). OAC, however, hinges on accurate channel-gain precoding and strict synchronization among edge devices, which are challenging in practice.
As such, how to design the maximum likelihood (ML) estimator in the presence of residual channel-gain mismatch and asynchronies is an open problem.
To fill this gap, this paper formulates the problem of misaligned OAC for FEEL and puts forth a whitened matched filtering and sampling scheme to obtain oversampled but independent, samples from the misaligned and overlapped signals.
Given the whitened samples, a sum-product ML (SP-ML) estimator and an aligned-sample estimator are devised to estimate the arithmetic sum of the transmitted symbols.
In particular, the computational complexity of our SP-ML estimator is linear in the packet length, and hence is significantly lower than the conventional ML estimator.
Extensive simulations on the test accuracy versus the average received energy per symbol to noise power spectral density ratio (EsN0) yield two main results:
1) In the low EsN0 regime, the aligned-sample estimator can achieve superior test accuracy provided that the phase misalignment is non-severe. In contrast, the ML estimator does not work well due to the error propagation and noise enhancement in the estimation process.
2) In the high EsN0 regime, the ML estimator attains the optimal learning performance regardless of the severity of phase misalignment. On the other hand, the aligned-sample estimator suffers from a test-accuracy loss caused by phase misalignment.
\end{abstract}

\begin{IEEEkeywords}
Federated edge learning, over-the-air computations, asynchronous, maximum likelihood estimation, sum product algorithm.
\end{IEEEkeywords}

\section{Introduction}
With the increasing adoption of Internet of Things (IoT) devices and services, exponentially growing amount of data is collected at the wireless network edge. Increasingly complex machine learning models are trained and deployed to gather intelligence from the data collected by edge devices \cite{NatureDL, Gunduz:JSAC:19}. While this is conventionally done at a cloud server \cite{Significant2020}, offloading huge amounts of edge data to centralized cloud servers is not sustainable, and will potentially cause significant network congestion \cite{Gunduz:CL:20}. Moreover, data from edge devices contain user-specific features, and centralized processing also causes privacy concerns.
Federated learning (FL) has been proposed as an alternative distributed solution to enable collaborative on-device learning without sharing private training data \cite{FL1, FL2, FL3}.

FL is an iterative distributed learning algorithm. In its basic implementation, orchestrated by a parameter server (PS), each iteration of FL consists of four main steps \cite{FL2}:
1) Downlink (DL) broadcast -- a PS maintains a global model and periodically broadcasts the latest global model to the edge devices;
2) Local training -- upon receiving the latest global model, each edge device trains the model locally using its local data set;
3) Uplink (UL) model aggregation -- after training, all, or a subset, of devices transmit their model updates back to the PS;
4) Global model update -- the PS updates the global model using the model updates collected from the edge devices, typically by taking their average.

In the case of edge devices, often the devices that collaborate to learn a common model are within physical proximity of each other, and are coordinated by a nearby access point, e.g., a base station acting as the PS. In this, so-called federated edge learning (FEEL) scenario \cite{Gunduz:CL:20}, the UL model aggregation step is particularly challenging as the wireless medium is shared among all the participating devices. Traditional radio access network (RAN) technologies distribute channel resources among the devices by means of orthogonal multiple-access technologies \cite{gupta2015survey} (e.g., TDMA, CDMA, OFDMA).
However, such orthogonal resource allocation techniques significantly limit the quality of model updates that can be sent from individual devices due to the limited channel resource that can be allocated to each device. We note that the number of real values to be transmitted by each device scales according to the neural network size. For today's neural networks, this number can easily run into hundreds of millions or more \cite{he2016deep}, and hence, is a heavy burden for the RANs.

Analog over-the-air computation (OAC) is a promising technique to realize  uplink model aggregation in an efficient manner \cite{Gunduz1, Gunduz2, AirComp:Huang, sery2020over, ShanghaiTech, Zhu:TWC:21, Amiri:GlobalSIP19, amiri2020blind, Sun:ICC:20}. The basic idea of OAC is to create and leverage inter-user interferences over the multiple-access channel (MAC) rather than trying to avoid it. When operated with OAC, devices send their model updates in an uncoded fashion by directly mapping each model parameter to a channel symbol: each device first precodes the transmitted symbols by the inverse of the UL channel gain (assumed to be known to the transmitter in advance) and then transmits the precoded symbols to the receiver in an analog fashion. All the participating devices transmit simultaneously in the same communication link such that their signals overlap at the PS. Provided that the channel-gain precoding and transmission timing are accurate, the fading MAC reduces to a Gaussian MAC and the  signal overlapped from the devices to the PS over-the-air naturally produces the arithmetic sum of the local model-updates \cite{amiri2020blind}.

Compared with the traditional digital multiple-access sch\-emes, wherein the communication and computation constitute separate processes, OAC is a joint comput\-ation-and-com\-munica\-tion scheme exploiting the fact that the MAC inherently yields an additive superposed signal.

The successful operation of OAC hinges on accurate channel-gain precoding and strict synchronization among the participating devices \cite{Gunduz2, AirComp:Huang}. In practice, however, both requirements may not be perfectly fulfilled.
On the one hand, the channel-gain precoding at the edge devices can be imperfect due to the inaccurate channel estimation and non-ideal hardware.
The consequence is that there can be residual channel-gain mismatch in the overlapped signals.
On the other hand, to meet the synchronization requirement, each device has to carefully calibrate the transmission timing -- based on its distance from the PS and its moving speed -- so that their signals overlap exactly with each other at the PS. This strict synchronization across different devices is very expensive to realize in practice, and there can be residual asynchronies among the signals from different devices.

With the residual channel gains and residual asynchronies in the system, which we refer to as the {\it misaligned OAC}, an open problem is how to estimate the arithmetic sum of the transmitted symbols from different devices.
This paper fills this gap and addresses the key problem in the misaligned OAC on how to devise the maximum likelihood (ML) estimator in the face of the channel-gain and time misalignments among signals.

Our main contributions are as follows:
\begin{enumerate}[leftmargin=0.65cm]
\item We formulate the problem of misaligned OAC for FEEL considering a time-domain realization of OAC. We put forth a whitened matched filtering and sampling (WMFS) scheme that yields oversampled, but independent, samples from the overlapped signals. An ML estimator for the arithmetic sum based on the whitened samples is devised.
\item To tackle the inter-symbol and inter-device interferences in the misaligned OAC, ML estimation requires the inversion of a large coefficient matrix, and hence, is computationally intensive. In view of this, we dissect the inner structure of the whitened samples and put forth a factor-graph based ML estimator exploiting the sparsity of the coefficient matrix. This factor-graph estimator, dubbed sum-product ML (SP-ML) estimator, interprets the compositions of samples by a factor graph and computes the likelihood functions via an analog message passing process on the graph. With the SP-ML estimator, the computational complexity of ML estimation is significantly reduced from $\Omega(L^2\log L)$ to $\Omega(L)$ for a packet of length $L$.
\item We identify two main problems of ML estimation in the misaligned OAC: error propagation and noise enhancement. As a result, ML estimation does not work well in the low average received energy per symbol to noise power spectral density ratio (EsN0) regime. To tackle this problem, we further put forth an aligned-sample estimator for the misaligned OAC leveraging a subsequence of whitened samples, wherein the symbols from different devices are ``aligned'', i.e., the indexes of symbols from different devices are consistent in these samples. This estimator is shown to be a good alternative to the ML estimator in the low-EsN0 regime. The complexity of the aligned-sample estimator is also linear in the packet length.
\item With the ML and aligned-sample estimators for the misaligned OAC, we perform extensive simulations on the CIFAR dataset varying the degrees of time misalignment, phase misalignment, and EsN0. The learning performance is measured by means of test accuracy, i.e., the achieved accuracy of the learned neural network on the test data set.
We find that
i) When there is no phase misalignment, the test accuracies of the ML estimator and the aligned-sample estimator are on the same footing for various degrees of time misalignment.
ii) When there is phase misalignment, the ML estimator works only in the high-EsN0 regime whereas the aligned-sample estimator works in both the low and high EsN0 regimes. Nevertheless, the aligned-sample estimator suffers from a loss in the test accuracy due to phase misalignment. In particular, the larger the phase misalignment, the greater the test-accuracy loss. In the case of severe phase misalignment, the aligned-sample estimator leads to learning divergence even in the noiseless case.
The ML estimator, on the other hand, does not incur such test-accuracy loss even with severe phase misalignment.
iii) When there is phase misalignment, the ML estimator benefits from time asynchronicity while the aligned-sample estimator suffers from time asynchronicity. iv) Overall, the aligned-sample estimator is preferred in the low-EsN0 regime and the ML estimator is preferred in the high-EsN0 regime.
\end{enumerate}

\textbf{Notations} -- We use boldface lowercase letters to denote column vectors (e.g., $\bm{\theta}$, $\bm{s}$) and boldface uppercase letters to denote matrices (e.g., $\bm{V}$, $\bm{D}$). For a vector or matrix, $(\cdot)^\top$ denotes the transpose, $(\cdot)^*$ denotes the complex conjugate, $(\cdot)^H$ denotes the conjugate transpose, and $(\cdot)^\dagger$ denotes the Moore-Penrose pseudoinverse. $\mathbb{R}$ and $\mathbb{C}$ stand for the sets of real and complex numbers, respectively. $(\cdot)^\mathfrak{r}$ and $(\cdot)^\mathfrak{i}$ stand for the real and imaginary components of a complex symbol or vector, respectively. The imaginary unit is represented by $j$. $\mathcal{N}$ and $\mathcal{CN}$ stand for the real and complex Gaussian distributions, respectively. The cardinality of a set $\mathcal{V}$ is denoted by $|\mathcal{V}|$. The sign function is denoted by $\text{sgn}(\cdot)$.

\section{System Model}\label{sec:II}
\begin{figure}[t]
  \centering
  \includegraphics[width=0.8\columnwidth]{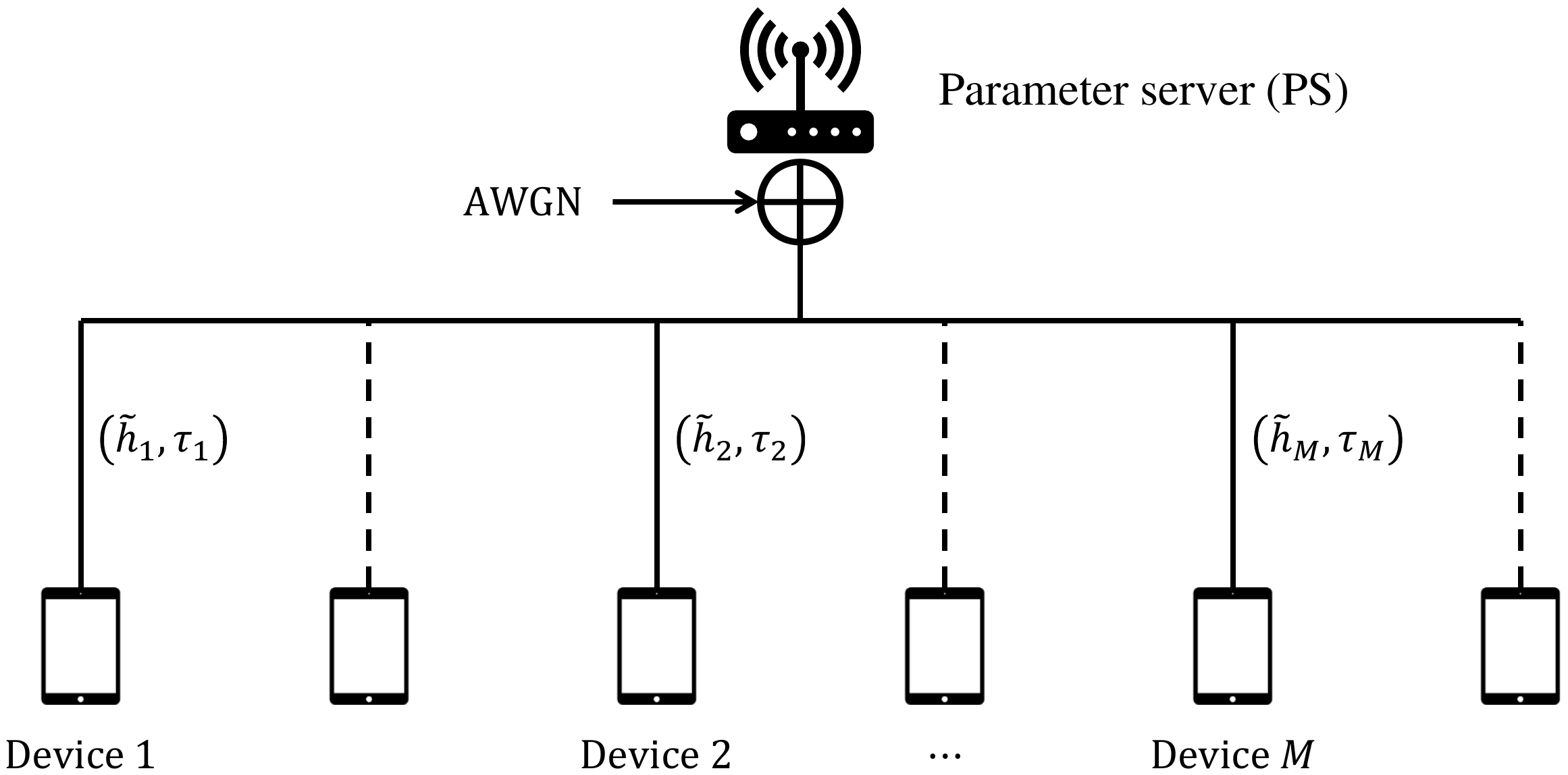}\\
  \caption{In FEEL, edge devices collaboratively train a shared model with the help of a wireless access point acting as a PS.}
\label{fig:1}
\end{figure}

We consider FEEL with the help of a wireless PS where nearby edge devices with distinct local datasets collaborate over the shared wireless medium to train a common model, as shown in Fig.~\ref{fig:1}. The learning process goes through many iterations. Without loss of generality, let us focus on one of the iterations, wherein $M$ active devices participate in the training. The iteration proceeds as follows~\cite{FL2}:

\begin{enumerate}[leftmargin=0.6cm]
\item DL broadcast: at the beginning of the iteration, the PS broadcasts the global model $\bm{\theta}\in\mathbb{R}^d$ to the $M$ edge devices;
\item Local training: each of the $M$ devices trains the global model $\bm{\theta}$ on its local dataset $\mathcal{B}_m$ of size $B_m$ and obtains a new model $\tilde{\bm{\theta}}_m \in\mathbb{R}^d$;
\item UL aggregation: each device scales the local model update $\tilde{\bm{\theta}}_m-\bm{\theta}$ by $B_m$ and transmits the scaled model update $\bm{\theta_m^\prime} =B_m (\tilde{\bm{\theta}}_m-\bm{\theta})\in\mathbb{R}^d$ back to the PS;
\item Arithmetic-sum estimation: the PS estimates the arithmetic sum of the transmitted model-updates $\bm{\theta_m^\prime}$ from the edge devices:
\begin{eqnarray}\label{eq:II1}
\bm{\theta}_+=\sum_{m=1}^{M}\bm{\theta_m^\prime};
\end{eqnarray}
\item Model update: the PS updates the global model by
\begin{eqnarray}\label{eq:II2}
\bm{\theta}_\text{new}=\bm{\theta}+\frac{1}{\sum_m B_m}\bm{\theta}_+.
\end{eqnarray}
\end{enumerate}

The updated global model $\bm{\theta}_\text{new}$ is then broadcasted in the next iteration and the cycle continues.

\begin{rem}
To compute \eqref{eq:II2}, the sum of the dataset sizes $\sum_m B_m$  has to be known to the PS a priori. Thus, we let each device transmit the local dataset size $B_m$ to the PS reliably in advance of data transmission (in a digital manner over a control channel, with channel coding and automatic repeat request, for example).
\end{rem}

Among the above five steps, the uplink model aggregation poses the greatest challenge to the RAN. In this step, each device has to transmit $d$ real numbers to the PS, where $d$ can run into hundreds of millions or more.
We consider analog OAC to realize the uplink model aggregation in this paper.

When operated with OAC, edge devices transmit their raw model updates $\bm{\theta_m^\prime}$ simultaneously to the PS in an analog manner (without digital modulation and channel coding). The PS, on the other hand, estimates the sum of the model updates $\bm{\theta_+}$ directly from the received overlapped signal.

The signal flow is detailed as follows. Each device partitions its sequence of scaled model update $\bm{\theta_m^\prime}\in\mathbb{R}^d$ to two subsequences $\bm{\theta_m^\prime}=\allowbreak[(\bm{s_m^\mathfrak{r}})^\top,\allowbreak(\bm{s_m^\mathfrak{i}})^\top]^\top$, where $\bm{s_m^\mathfrak{r}}$, $\bm{s_m^\mathfrak{i}}\in\mathbb{R}^{d/2}$, and constructs a complex sequence $\bm{s_m}\in\mathbb{C}^{d/2}$: $\bm{s_m}=\bm{s_m^\mathfrak{r}}+j\bm{s_m^\mathfrak{i}}$,
that is, the raw model update information $\bm{\theta_m^\prime}$ is carried on both the real and imaginary parts of $\bm{s_m}$.

Time is divided into slots, and each device transmits a packet of $L$ symbols in each slot where $\bm{s_m}\allowbreak=[\allowbreak s_m[1],\allowbreak s_m[2],\allowbreak ...,\allowbreak s_m[L]]^\top$. Therefore, to transmit the total number of $d/2$ complex symbols, $\lceil d/2L\rceil$ slots are needed. Without loss of generality, we focus on the signal processing in one slot.

The time-domain signal transmitted by the $m$-th device in one slot is given by
\begin{eqnarray}\label{eq:II3}
x_m(t)=\alpha_m \sum_{\ell=1}^L s_m[\ell]p(t-\ell T),
\end{eqnarray}
where 1) $p(t)=1/2 \left[\text{sgn}(t+T)-\text{sgn}(t)\right]$ is a rectangular pulse of duration $T$; 2) $\alpha_m$ is the channel precoding factor.
Given an estimated channel coefficient $\bar{h}_m$ at the $m$-th device, $\alpha_m$ is designed to be $\alpha_m= 1/\bar{h}_m$.\footnote{In practice, the channel-gain precoding is limited by the maximum transmission power of the edge devices.
In the case of deep fading $\alpha_m$ would be very large, and we would have to clip $\alpha_m$ to satisfy the peak or average transmission-power constraint. In our formulation, this may be one cause of the residual channel gains at the receiver.}
Each of the $M$ edge devices then  calibrates the transmission timing, based on its distance from the PS and its moving speed, so that the signals from
different devices arrive at the PS simultaneously.

\begin{figure}[t]
  \centering
  \includegraphics[width=0.85\columnwidth]{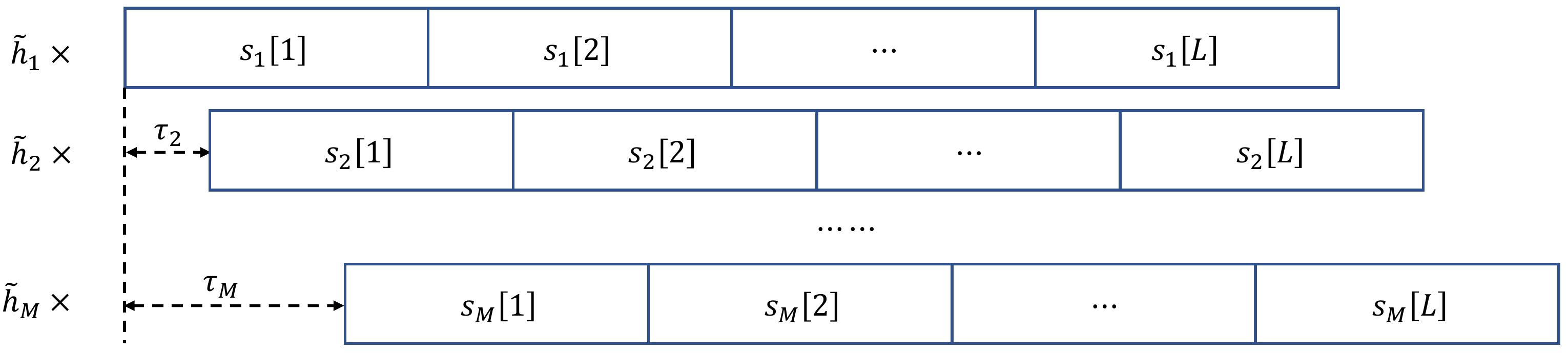}\\
  \caption{In each slot, the transmitted packets from different devices overlap at the PS with channel misalignments and relative time offsets.}
\label{fig:2}
\end{figure}

In practice, however, both the channel-gain precoding and tr\-ansmission-timing calibration can be imperfect due to the non-ideal hardware and inaccurate estimation of the channel gains and transmission delays.
The received signal $r(t)$ at the PS can be written as
\begin{eqnarray}\label{eq:II4}
r(t)=\sum_{m=1}^M \tilde{h}_m x_m(t-\tau_m) +z(t),
\end{eqnarray}
where
\begin{enumerate}[leftmargin=0.6cm]
\item $\tilde{h}_m$ is the time-domain complex channel response. We consider flat fading (frequency nonselective) and slow fading (time nonselective) channels \cite{TseBook}. That is, for the channel between each device and the PS, the maximum delay spread is less than the symbol period $T$ so that there is only one resolvable path with the channel response $\tilde{h}_m$ at the receiver, and $\tilde{h}_m$ remains constant over one packet.\footnote{To ease exposition, the main body of this paper considers only the channel-gain misalignment, time misalignment, and slow fading channel. However, the system model can be easily generalized to OAC with residual carrier frequency offset (CFO) and fast fading channel, as detailed in Appendix \ref{sec:AppA}.}
\item Without loss of generality, we sort the $M$ devices so that the symbols from the devices with smaller indexes arrive at the receiver earlier. The delay of the first device is set to $\tau_1=0$, and the relative delay of the $m$-th device with respect to the first device is denoted by $\tau_m$. We assume the time offsets $\tau_m$, $\forall~m$, are less than the symbol duration $T$, as shown in Fig.~\ref{fig:2}.
In the ideal case where the timing calibrations are perfect, the relative delays among packets are $\tau_m=0$, $\forall m$.
\item $z(t)$ is the zero-mean baseband complex additive white Gaussian noise (AWGN), the double-sided power spectral densities of each of its real and imaginary parts is $N_0/2$ for an aggregate of $N_0$.
\end{enumerate}

Substituting \eqref{eq:II3} into \eqref{eq:II4} gives us
\begin{eqnarray}\label{eq:rt}
r(t)\hspace{-0.2cm}&=&\hspace{-0.2cm}\sum_{m=1}^M\tilde{h}_m \alpha_m\sum_{\ell=1}^L s_m[\ell]p(t-\tau_m-\ell T) + z(t) \nonumber\\
\hspace{-0.2cm}&=&\hspace{-0.2cm} \sum_{\ell=1}^L \sum_{m=1}^M {h}'_m  s_m[\ell]p(t-\tau_m-\ell T) + z(t),
\end{eqnarray}
where $h'_m=\tilde{h}_m/\bar{h}_m$ is the residual channel-fading coefficient between the $m$-th device and the PS.

Succinctly speaking, there can be two kinds of misalignments among the signals transmitted from different devices: 1) channel-gain misalignment ${h}'_m$ caused by inaccurate channel-gain precoding; and 2) time misalignment $\tau_m$ caused by inaccurate calibration of the transmission timing.

The objective of the PS is to estimate $\bm{\theta}_+=\sum_{m=1}^{M}\bm{\theta_m^\prime}$, i.e., the arithmetic sum of the local model updates. This is equivalent to estimating the arithmetic sum of the transmitted complex symbols $\bm{s_+}$, where
\begin{eqnarray*}
s_+[i] \triangleq \sum_{m=1}^{M}s_m[i],
\end{eqnarray*}
because $\bm{\theta_m^\prime}$ are carried on the real and imaginary parts of the $\bm{s_m}$ sequence. In other words, given the estimated sequence $\bm{\hat{s}}_+$, the estimated arithmetic sum of the local model updates is $\bm{\hat{\theta}}_+ = \left[(\bm{\hat{s}}_+^\mathfrak{r})^\top,(\bm{\hat{s}}_+^\mathfrak{i})^\top\right]^\top$.
Therefore, we shall focus on the estimation of the complex symbols $\bm{s_+}$ in the following sections.

\begin{rem}
This paper formulates misaligned OAC for FEEL considering a time-domain realization. OAC can also be realized in the frequency domain via OFDM. The connections and differences between the two realizations are discussed in the Conclusion section.
\end{rem}

\section{Aligned and Misaligned OAC}\label{sec:III}
\subsection{Aligned OAC}
Prior works on OAC, with the exception of \cite{amiri2020blind}, considered only the perfectly aligned case  \cite{Zhu,Gunduz1,Gunduz2,ShanghaiTech,sery2020over}, where there is neither channel-gain misalignment nor time misalignment, which we refer to as the {\it aligned OAC}.
In this case, we have $\alpha_m=1/\tilde{h}_m$ and $\tau_m=0$, $\forall m$, and the received signal is given by
\begin{eqnarray}\label{eq:II6}
r(t)=\sum_{\ell=1}^L\sum_{m=1}^M s_m[\ell]p(t-\ell T) +z(t).
\end{eqnarray}

Matched filtering $r(t)$ by the same rectangular pulse $p(t)$ and sampling at $t=iT$, $i=1,2,...,L$, gives us
\begin{eqnarray}\label{eq:II7}
r[i]\!=\!\frac{1}{T}\int_{(i-1)T}^{iT}\!\!r(t) dt = \!\sum_{m=1}^{M}\! s_m[i] \!+\! z[i] = s_+[i] \!+\! z[i],
\end{eqnarray}
where the noise sequence $z[i]$ in the samples is independent and identically distributed (i.i.d.), $z[i]\sim\mathcal{CN}(0,\frac{N_0}{T})$.

As can be seen, the target signal $s_+[i]$ appears explicitly on the right hand side (RHS) of \eqref{eq:II7}. In this context, the fading MAC degenerates to a Gaussian MAC and the $M$ devices can be abstracted as a single device transmitting the arithmetic sum of the local model updates directly to the PS.
In practice, however, the channel-gain precoding and the calibration of transmission timing can be inaccurate. With either channel-gain or time misalignment, clean samples as in \eqref{eq:II7} with $\bm{s}_+$ explicitly present are no longer available.

\subsection{Misaligned OAC}
With channel-gain and time misalignments, the received signal $r(t)$ is given in \eqref{eq:rt} and illustrated in Fig.~\ref{fig:2}.
Let us first follow the standard signal processing flow in digital communications to process the received signal. Specifically, we first matched filter $r(t)$ by the rectangular pulse $p(t)$ and then oversample the matched filtered signal at $\{iT+\tau_k: i=1,2, ...,L; k=1,2,...,M\}$ to collect sufficient statistics \cite{VerduBook}. In so doing, the samples we get, denoted by $\{r_k[i]:k=1,2,...,M;i=1,2,...,L\}$, can be written as
\begin{eqnarray}\label{eq:sample1}
\hspace{-0.65cm}&& r_k[i]=\frac{1}{T}\int_{(i-1)T+\tau_k}^{iT+\tau_k}\hspace{-0.3cm} r(t)\ast p(t) \,dt  \\
\hspace{-0.65cm}&& =\frac{1}{T} \sum_{m=1}^M  \int_{(i-1)T+\tau_{k}}^{(i-\mathbbm{1}_{m>k})T+\tau_{m}}\hspace{-0.3cm} h'_m s_m[i\!-\!\mathbbm{1}_{m>k}]\,d\zeta + \frac{1}{T}\sum_{m=1}^M   \nonumber\\
\hspace{-0.65cm}&& \qquad \int_{(i-\mathbbm{1}_{m>k})T+\tau_{m}}^{iT+\tau_{k}}\hspace{-0.4cm} h'_m s_m[i\!+\!\mathbbm{1}_{m<k}]\,d\zeta \!+\!\frac{1}{T} \!\int_{(i-1)T+\tau_{k}}^{iT+\tau_{k}} \hspace{-0.3cm}z(\zeta)\,d\zeta  \nonumber\\
\hspace{-0.65cm}&& \triangleq \sum_{m=1}^M \!c_{m,k}[i] s_m[i\!-\!\mathbbm{1}_{m>k}] \!+\! \sum_{m=1}^M\! c^\prime_{m,k}[i] s_m[i\!+\!\mathbbm{1}_{m<k}] \!+\! z_k[i],\nonumber
\end{eqnarray}
where $\mathbbm{1}$ is the indicator function and
\begin{eqnarray*}
c_{m,k}[i] \hspace{-0.2cm}&=&\hspace{-0.2cm} \frac{h^\prime_m}{T} [(1-\mathbbm{1}_{m>k})T+\tau_{m} - \tau_{k}],\\
c^\prime_{m,k}[i] \hspace{-0.2cm}&=&\hspace{-0.2cm} \frac{h^\prime_m}{T} [\mathbbm{1}_{m>k}T + \tau_{k} - \tau_{m}].
\end{eqnarray*}
As shown in Appendix \ref{sec:AppA}, when there is residual CFO and the channel is fast fading, the discrete samples can be written in the same form as \eqref{eq:sample1} with the coefficients $c_{m,k}[i]$ and $c^\prime_{m,k}[i]$ given in \eqref{eq:A1} and \eqref{eq:A2}.

The noise sequence $\{z_k[i]\}$ in \eqref{eq:sample1} is colored since
\begin{eqnarray}\label{eq:samplesNoise}
\hspace{-0.65cm}&& \mathbb{E}[z_{k}[i]z_{k^\prime}[i^\prime]]=\frac{1}{T^2}\! \int_{(i-1)T+\tau_{k}}^{iT+\tau_{k}} \int_{(i^\prime-1)T+\tau_{k^\prime}}^{i^\prime T+\tau_{k^\prime}} \hspace{-0.3cm}z(\zeta)z(\zeta^\prime)\,d\zeta\,d\zeta^\prime \nonumber\\
\hspace{-0.65cm}&& = \begin{cases}
\frac{N_0}{T^2}\lambda(i,i^\prime,k,k^\prime) & \text{if}~ \lambda(i,i^\prime,k,k^\prime)\in[0,T), \\
\frac{N_0}{T^2}[2T\!-\!\lambda(i,i^\prime,k,k^\prime)] & \text{if}~ \lambda(i,i^\prime,k,k^\prime)\in[T,2T), \\
0 & \text{otherwise},
\end{cases}
\end{eqnarray}
where $\lambda(i,i^\prime,k,k^\prime) = (i^\prime\!-\!i\!-\!1) T\!+\!\tau_{k^\prime} \!-\! \tau_{k}$.

Given the samples $\{r_k[i]\}$ in \eqref{eq:sample1}, we now set out to estimate the desired arithmetic sum $\bm{s_+}$. First, the sequence of samples $y_k[i]$ can be written in a more compact form as
\begin{eqnarray}\label{eq:samples1Mat}
\bm{r}= \bm{As+z},
\end{eqnarray}
where the dimensionalities of $\bm{r}$, $\bm{s}$, and $\bm{z}$ are $ML\times 1$, and the dimensionality of matrix $\bm{A}$ is $ML\times ML$ (see the detailed form of \eqref{eq:samples1Mat} at the top of the next page). Denoting by $\Sigma_z$ the covariance matrix of the noise sequence $\bm{z}$, each element of $\Sigma_z$ can then be computed from \eqref{eq:samplesNoise}.

\begin{figure*}
\begin{eqnarray*}
\begin{bmatrix}
\begin{smallmatrix}
r_1[1] \\
...    \\
r_M[1] \\
r_1[2] \\
...    \\
r_M[2] \\
...\\
r_1[L] \\
...    \\
r_M[L] \\
\end{smallmatrix}
\end{bmatrix}\!\!=\!\!
\begin{bmatrix}
\begin{smallmatrix}
c_{1,1}[1] & c^\prime_{2,1}[1] & c^\prime_{3,1}[1] & ...   &  c^\prime_{M,1}[1] &             &               &     &               &    &   &    &  \\
c_{1,2}[1] & c_{2,2}[1] & c^\prime_{3,2}[1] & ...   &  c^\prime_{M,2}[1] & c^\prime_{1,2}[2]  &               &     &               &    &   &    &  \\
...        & ...        & ...        & ...   &  ...        &  ...        & ...           &     &               &    &   &    &  \\
c_{1,M}[1] & c_{2,M}[1] & c_{3,M}[1] & ...   &  c_{M,M}[1] & c^\prime_{1,M}[2]  & c^\prime_{2,M}[2]    & ... & c^\prime_{M\!-\!1,M}[2]  &    &   &    &  \\
           & c_{2,1}[1] & c_{3,1}[1] & ...   &  c_{M,1}[1] & c_{1,1}[2]  & c^\prime_{2,1}[2]    & c^\prime_{3,1}[2]   &  ... & c^\prime_{M,1}[2] &    &  &    \\
           &            & c_{3,2}[1] & ...   &  c_{M,2}[1] & c_{1,2}[2]  & c_{2,2}[2]    & c^\prime_{3,2}[2]   &  ... & c^\prime_{M,2}[2] & c^\prime_{1,2}[3] & &  \\
           &            & ...        & ...   &  ...        & ...         & ...           & ...  & ...         & ... & ... & &    \\
           &            &            &       &  c_{1,M}[2] & c_{2,M}[2]  & c_{3,M}[2]    & ...  & c_{M,M}[2]  & c^\prime_{1,M}[3] & c^\prime_{2,M}[3] & ...& c^\prime_{M\!-\!1,M}[3]    \\
           &            &            &       &             & ...         &  ...          & ...  &  ...        & ...        & ...        & ...& ... & ... &
\end{smallmatrix}
\end{bmatrix}
\!\!
\begin{bmatrix}
\begin{smallmatrix}
s_1[1] \\
...    \\
s_M[1] \\
s_1[2] \\
...    \\
s_M[2] \\
...\\
s_1[L] \\
...    \\
s_M[L] \\
\end{smallmatrix}
\end{bmatrix}
\!\!+\!\!
\begin{bmatrix}
\begin{smallmatrix}
z_1[1] \\
...    \\
z_M[1] \\
z_1[2] \\
...    \\
z_M[2] \\
...\\
z_1[L] \\
...    \\
z_M[L] \\
\end{smallmatrix}
\end{bmatrix}
\end{eqnarray*}
\hrulefill
\end{figure*}

The desired sequence $\bm{s_+}$, on the other hand, can be written as a linear transformation of the complex vector $\bm{s}$:
\begin{eqnarray}\label{eq:s_plus_matrix}
\bm{s_+}= \bm{V}\bm{s},
\end{eqnarray}
where the $L\times ML$ matrix $\bm{V}$ is given by
\begin{eqnarray*}
\bm{V} = \begin{bmatrix}
\bm{1}_{1\times M} &                    &                    &                    \\
                   & \bm{1}_{1\times M} &                    &                    \\
                   &                    & ...                &                    \\
                   &                    &                    & \bm{1}_{1\times M}
\end{bmatrix},
\end{eqnarray*}
in which $\bm{1}_{1\times M}$ represents a $1\times M$ all-ones matrix.

Multiplying both sides of \eqref{eq:samples1Mat} by $\bm{VA^{-1}}$ gives us
\begin{eqnarray}\label{eq:processedSample1}
\bm{VA^{-1}}\bm{r}= \bm{s_+}+\bm{VA^{-1}z},
\end{eqnarray}
based on which an ML estimator can be devised, as in Definition \ref{defi:2}.

\begin{defi}[ML estimation for misaligned OAC]\label{defi:2}
Given a sequence of samples $\bm{r}\in\mathcal{C}^{ML}$ in \eqref{eq:samples1Mat}, the ML estimate of sequence $\bm{s}_+\in\mathcal{C}^L$ is
\begin{eqnarray}\label{eq:ML1}
\hat{\bm{s}}^\text{ml}_+ = \bm{VA^{-1}}\bm{r}.
\end{eqnarray}
\end{defi}

Eq. \eqref{eq:ML1} follows directly from \eqref{eq:processedSample1} since the likelihood function of $\bm{s_+}$ is an $L$-dimensional Gaussian distribution. Specifically, given $\bm{r}$, the likelihood function of $\bm{s_+}$ is
\begin{eqnarray*}
f(\bm{VA^{-1}}\bm{r}|\bm{s_+}) \propto \mathcal{CN}(\bm{VA^{-1}}\bm{r}, \bm{VA^{-1}}\Sigma_z\bm{A^{-H}V^H}).
\end{eqnarray*}
Differentiating $f(\bm{VA^{-1}}\bm{r}|\bm{s_+})$ with respect to $\bm{s_+}$ gives us the ML estimate $\hat{\bm{s}}^\text{ml}_+$ in \eqref{eq:ML1}.

An important implication of \eqref{eq:ML1} is that, the maximum-likelihood $\bm{s_+}$ can be obtained by first finding the maximum-likelihood transmitted vector $\hat{\bm{s}}^\text{ml}=\bm{A^{-1}r},$
and then performing the arithmetic sum
$\hat{\bm{s}}^\text{ml}_+ = \bm{V}\hat{\bm{s}}^\text{ml}$. Said in another way, the ML estimation of $\bm{s_+}$ boils down to multi-user estimation/detection (MUE) {when we have misaligned channel gains at the receiver}.

\begin{rem}
The transmitted vector $\bm{s}$ carries the local weight-updates of a neural network. Therefore, $\bm{s}$ is by no means i.i.d. considering the strong correlations among the weights of the neural network. This also suggests that the prior information of $\bm{s}$ is hard to obtain and is unlikely to be known to the PS in advance. As a result, ML estimation is the only choice at the receiver.
\end{rem}

\begin{rem}
{The result that the ML estimation of $\bm{s_+}$ boils down to MUE is exclusive to the misaligned OAC system, wherein the transmitted symbols are continuous valued and the channel gains are misaligned.}
In digital communications, we do not have such result for ML estimation. The reason for this divergence is as follows.
OAC is an analog communication system wherein the transmitted symbols $\bm{s}$ are continuous complex values.
In contrast, the transmitted symbols in digital communications are discrete constellations.
Whether the prior probability distribution of the transmitted symbols is available to the receiver or not, the constellation itself serves as a kind of prior information as the detection space is naturally narrowed down to the possible constellation points.
As a result, when we perform ML estimation in digital communications, we inherently assume all the constellations are equiprobable.
In that case, the likelihood function $f(\bm{r}|\bm{s})$ is a Gaussian mixture instead of Gaussian and the MUE-and-sum estimation is no longer optimal if we were to estimate the arithmetic sum $\bm{s}_+$.

On the other hand, when we perform ML estimation in OAC, all the complex space is assumed to be equiprobable. The only information we have is the noise-contaminated sample and the likelihood function is a Gaussian centered around the noisy sample.
Two implications about the ML estimation in OAC are thus 1) it faces an infinitely large estimation space;
2) it can be very susceptible to noise.
\end{rem}

The ML estimator in \eqref{eq:ML1} is not a practical estimator due to the prohibitive computational complexity of matrix inversion. To invert an $n$ by $n$ matrix, the best proven lower bound of the computational complexity is $\Omega(n^2\log n)$ \cite{complexity}.
Notice that the dimensionality of $\bm{A}$ is $ML$ by $ML$. Thus, the computational complexity of \eqref{eq:ML1} is $\Omega(L^2M^2\log(LM))$.
In practical OAC systems, $M$ cannot be too large due to the saturation of the receiver (that is, the received signal power can exceed the dynamic range of the receiver if $M$ is too large), but the packet length $L$ can be extremely large.
Let us fix $M$ as a constant, the computational complexity of \eqref{eq:ML1} is then $\Omega(L^2\log L)$.

To address this problem and devise an ML estimator with acceptable computational complexity, we put forth in Section~\ref{sec:IV} a factor-graph based ML estimator by exploiting the sparsity of the coefficient matrix.
Compared with the ML estimator in \eqref{eq:ML1}, the computational complexity of the factor-graph based ML estimator is only $\Omega(L)$.

\section{A Sum-Product ML Estimator and The Aligned-sample Estimator }\label{sec:IV}
Before we dive deeper to dissect the inner structure of the coefficient matrix, let us first introduce a new matched filtering and sampling scheme that gives us oversampled but independent samples, which we refer to as the {\it whitened matched filtering and sampling} (WMFS).
Two benefits of the WMFS scheme are
1) the independent samples obtained from the scheme allows us to construct a factor graph with a simple structure, based on which a low-complexity SP-ML estimator can be devised;
2) the whitened scheme yields a subsequence of samples in which the indexes of symbols from different devices are consistent -- in these samples, the symbols from different devices are aligned in time, as shown in \eqref{eq:samplesMth}. This admits an aligned-sample estimator for the misaligned OAC.

\subsection{WMFS}\label{sec:IVA}
The key idea of the WMFS scheme is to use a bank of matched filters of different lengths to collect power judiciously from $r(t)$. Specifically, instead of using the rectangular pulse $p(t)$ as the matched filter as in \eqref{eq:sample1}, we define $M$ matched filters $\{p_k^\prime (t):k=1,2,...,M\}$ as follows:
\begin{eqnarray}\label{eq:matchedfilters}
p_k^\prime(t)=\frac{1}{2}\big[\text{sgn}(t+T)-\text{sgn}(t+T-d_k)  \big],
\end{eqnarray}
where the length of the $k$-th matched filter is $d_k=\tau_{k+1}-\tau_k$, $k=1,2,...,M$. For completeness, we define $\tau_{M+1}=T$.

\begin{figure}[t]
  \centering
  \includegraphics[width=1\columnwidth]{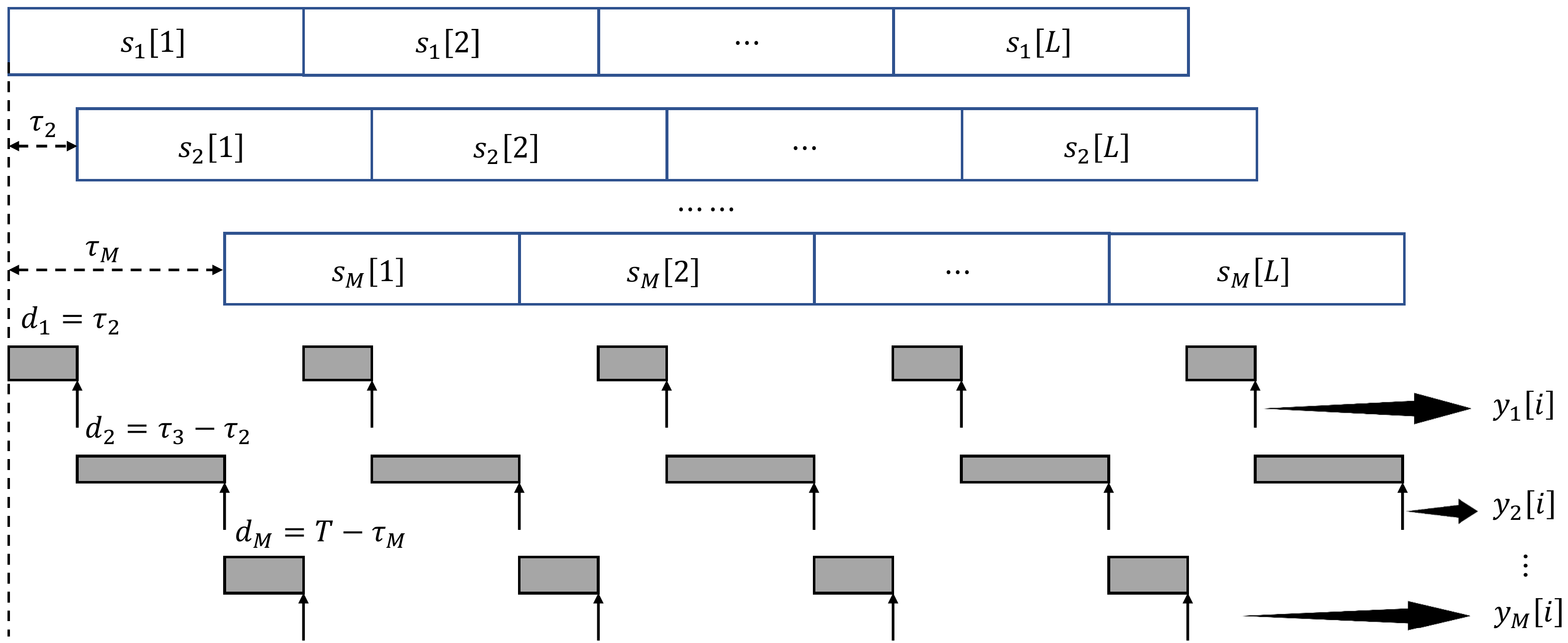}\\
  \caption{Matched filtering the received signal by a bank of $M$ filters of lengths $d_k=\tau_{k+1}-\tau_k$.}
\label{fig:3}
\end{figure}

The matched filtering and sampling processes are illustrated in Fig.~\ref{fig:3}. As shown, the signal filtered by the $k$-th matched filter is given by
\begin{eqnarray*}
y_k(t)=\frac{1}{d_k}\int_{-\infty}^{\infty}r(\zeta)p^\prime_k(t-\zeta) \,d\zeta,
\end{eqnarray*}
and we sample $y_k(t)$ at $(i-1)T\allowbreak+\tau_{k+1}:\allowbreak i=1,2,...,L+1$, giving
\begin{eqnarray}\label{eq:samples}
\hspace{-0.5cm}&& y_k[i] = y_k(t=(i-1)T+\tau_{k+1})=\frac{1}{d_k} \int_{(i-1)T+\tau_{k}}^{(i-1)T+\tau_{k+1}} \nonumber\\
\hspace{-0.5cm}&& \sum_{m=1}^M h'_m s_m[i-\mathbbm{1}_{m>k}]\,d\zeta +\frac{1}{d_k} \int_{(i-1)T+\tau_{k}}^{(i-1)T+\tau_{k+1}} \hspace{-0.3cm}z(\zeta)\,d\zeta  \nonumber\\
\hspace{-0.5cm}&& \triangleq \sum_{m=1}^M h'_m s_m[i-\mathbbm{1}_{m>k}] + \tilde{z}_k[i],
\end{eqnarray}
where we have defined $s_m[0]=s_m[L+1]=0$, $\forall m$, for completeness.

An important observation from \eqref{eq:samples} is that the noise term $\tilde{z}_k[i]\sim\mathcal{CN}(0, N_0/d_k)$ is independent for different $k$ and $i$. This is because
\begin{eqnarray}\label{eq:noiseVar}
\hspace{-0.7cm}&&\mathbb{E}\left[ \tilde{z}_k[i] \tilde{z}_{k^\prime}[i^\prime]\right] \nonumber\\
\hspace{-0.7cm}&& = \mathbb{E}\left[ \frac{1}{d_k} \int_{(i-1)T+\tau_{k}}^{(i-1)T+\tau_{k+1}} \!\!\!\tilde{z}(\zeta)\,d\zeta \frac{1}{d_{k^\prime}} \int_{(i^\prime-1)T+\tau_{k^\prime}}^{(i^\prime-1)T+\tau_{k^\prime+1}} \tilde{z}(\zeta^\prime)\,d\zeta^\prime  \right] \nonumber \\
\hspace{-0.7cm}&& = \frac{1}{d_kd_{k^\prime}} \int_{(i-1)T+\tau_{k}}^{(i-1)T+\tau_{k+1}} \!\!\!\!\int_{(i^\prime-1)T+\tau_{k^\prime}}^{(i^\prime-1)T+\tau_{k^\prime+1}}\!\!\!
\mathbb{E}\left[ \tilde{z}(\zeta)\tilde{z}(\zeta^\prime)\right]  \,d\zeta d\zeta^\prime \nonumber \\
\hspace{-0.7cm}&& = \frac{1}{d_kd_{k^\prime}} \int_{(i-1)T+\tau_{m}}^{(i-1)T+\tau_{m+1}} N_0\delta\left((i-i^\prime)(k-k^\prime)\right) \, d\zeta \nonumber\\
\hspace{-0.7cm}&& = \frac{N_0}{d_k} \delta\left((i-i^\prime)(k-k^\prime)\right).
\end{eqnarray}

Similarly to \eqref{eq:samples1Mat}, we can rewrite \eqref{eq:samples} in a matrix form as
\begin{eqnarray}\label{eq:samplesMat}
\bm{y}= \bm{Ds+\tilde{z}},
\end{eqnarray}
where the sequence of transmitted symbols $\bm{s}$ is the same as that in \eqref{eq:samples1Mat}. Unlike \eqref{eq:samples1Mat}, the vectors $\bm{y}$ and $\bm{\tilde{z}}$ in \eqref{eq:samplesMat} are $M(L+1)-1$ by $1$ dimensional, giving
\begin{small}
\begin{eqnarray*}
&&\hspace{-0.7cm}  \bm{y}=\Big[y_1[1],y_2[1],...,y_M[1],y_1[2],y_2[2],...,y_M[2],..., \\
&&\hspace{-0.7cm}  y_1[L],y_2[L],...,y_M[L], y_1[L\!+\!1],y_2[L\!+\!1],...,y_{M\!-\!1}[L\!+\!1] \Big]^\top, \\
&&\hspace{-0.7cm}  \bm{\tilde{z}}=\Big[\tilde{z}_1[1],\tilde{z}_2[1],...,\tilde{z}_M[1],\tilde{z}_1[2],\tilde{z}_2[2],...,\tilde{z}_M[2],..., \\
&&\hspace{-0.7cm}  \tilde{z}_1[L],\tilde{z}_2[L],...,\tilde{z}_M[L], \tilde{z}_1[L\!+\!1],\tilde{z}_2[L\!+\!1],...,\tilde{z}_{M\!-\!1}[L\!+\!1] \Big]^\top,
\end{eqnarray*}
\end{small}
and the coefficient matrix $\bm{D}$ is $M(L+1)-1$ by $ML$, giving
\begin{eqnarray}\label{eq:MatD}
\bm{D}=
\begin{bmatrix}
\begin{smallmatrix}
h_1^\prime &            &       &             &             &               &    &               &    &     \\
h_1^\prime & h_2^\prime &       &             &             &               &    &               &    &     \\
...        & h_2^\prime & ...   &             &             &               &    &               &     &   \\
h_1^\prime & ...        & ...   &  h_M^\prime &             &               &    &               &    &    \\
           & h_2^\prime & ...   &  h_M^\prime & h_1^\prime  &               &    &               &    &    \\
           &            & ...   &  ...        & h_1^\prime  & h_2^\prime    &    &               &     &   \\
           &            &       &  h_M^\prime & ...         & h_2^\prime    & ...  &             &    &    \\
           &            &       &             & h_1^\prime  & ...           & ...  & h_M^\prime  &    &    \\
           &            &       &             &             & h_2^\prime    & ...  & h_M^\prime  & ...&       \\
           &            &       &             &             &               & ...  & ...         & ... &   \\
           &            &       &             &             &               &      & h_M^\prime  & ... &     \\
           &            &       &             &             &               &      &             & ... &
\end{smallmatrix}
\end{bmatrix}.
\end{eqnarray}
When there is residual CFO and the channel is fast fading, matrix $\bm{D}$ can be written in the same form, as given in \eqref{eq:A3}.

The desired sequence at the PS is $\bm{s_+}= \bm{V}\bm{s}$, as in \eqref{eq:s_plus_matrix}.

Eq. \eqref{eq:noiseVar} validates that the noise sequence $\tilde{\bm{z}}$ is white, that is, \eqref{eq:samplesMat} can be viewed as a whitened model of \eqref{eq:samples1Mat}, and hence the name WMFS.

We emphasize that the signal model in \eqref{eq:samplesMat} is equivalent to that in \eqref{eq:samples1Mat} since they are sampled from the same received signal $r(t)$ and no information is lost.
More specifically, \eqref{eq:samplesMat} can be transformed back to \eqref{eq:samples1Mat} after some elementary row operations and row deletions.
Nevertheless, the model in \eqref{eq:samplesMat} is more favorable than \eqref{eq:samples1Mat} thanks to the following factors:
\begin{enumerate}
\item The whitened noise. As will be shown later, white noise admits a sample-by-sample factorization of the likelihood function and a much simpler structure of the factor graph.
\item The alleviated inter-symbol and inter-user interferences. A sample $y_k[i]$ is related to only $M$ complex symbols, each of which comes from a different device. In contrast, a sample $r_k[i]$ in \eqref{eq:samples1Mat} is related to $2M-1$ symbols.
\end{enumerate}

\begin{rem}
Since \eqref{eq:samples1Mat} and \eqref{eq:samplesMat} are equivalent, we can also design the ML estimator from \eqref{eq:samplesMat}.
Denoting the covariance matrix of $\bm{\tilde{z}}$ by $\bm{\Sigma_{\tilde{z}}}$ (it is a diagonal matrix since $\tilde{z}$ is white), the ML estimator is given by
\begin{eqnarray}\label{eq:ML2}
\hat{\bm{s}}^\text{ml}_+ = \bm{V}(\bm{D^H}\bm{\Sigma^{-1}_{\tilde{z}} D})^{-1}\bm{D^H\Sigma^{-1}_{\tilde{z}}y}.
\end{eqnarray}
Again, the inversion of $\bm{D^H}\bm{\Sigma^{-1}_{\tilde{z}} D}$ is computationally demanding, as in \eqref{eq:ML1}.
\end{rem}

\subsection{A Factor Graph Approach}
A possible way to reduce the complexity of the ML estimation is to exploit the sparsity of the coefficient matrix $\bm{D}$. To this end, let us focus on the ML estimate of a single entry in the desired sequence $\bm{s}_+$, i.e., $\hat{s}^\text{ml}_+[i]$.

Let $\bm{s[i]}=\big[s_1[i],s_2[i],...,s_M[i]\big]^\top$. Given an observed sample sequence $\bm{y}$, we have
\begin{eqnarray*}
\hat{s}^\text{ml}_+[i]\hspace{-0.2cm}&=&\hspace{-0.2cm}\arg\max_{s_+[i]}f(\bm{y}|\allowbreak s_+[i]) \\
\hspace{-0.2cm}&=&\hspace{-0.2cm}
\arg\max_{s_+[i]}
\int_{\bm{1}^\top\bm{s[i]}=s_+[i]} f(\bm{y}|\bm{s[i]})\,d\bm{s[i]}.
\end{eqnarray*}
In particular, $f(\bm{y}|\bm{s[i]})$ is a marginal function of $f(\bm{y}|\bm{s})$.
Thus, to find the ML estimate $\hat{s}^\text{ml}_+[i]$, a first step is to analyze $f(\bm{y}|\bm{s})$. In the following analysis, we will call $f(\bm{y}|\bm{s})$ the global likelihood function and $f(\bm{y}|\bm{s[i]})$ the marginal likelihood function.

In ML estimation, the transmitted symbols $\bm{s}$ are treated as constants. Randomness is only introduced by the noise sequence $\tilde{\bm{z}}$. Therefore, in the whitened model \eqref{eq:samplesMat}, the elements of $\bm{y}$ are independent of each other. We can then factorize the likelihood function $f(\bm{y}|\bm{s})$ as
\begin{equation}\label{eq:factorization}
f(\bm{y}|\bm{s}) \propto  \prod_{k=1}^M \prod_{i=1}^{L+1} f(y_k[i]|\bm{s}) \overset{(a)}{=} \prod_{k=1}^M \prod_{i=1}^{L+1} f(y_k[i]|\mathcal{V}(y_k[i])),\\
\end{equation}
where (a) follows because a sample $y_k[i]$ is related to only $M$ complex symbols in $\bm{s}$.
As per \eqref{eq:samples}, we denote these symbols by $\mathcal{V}(y_k [i])\allowbreak=\{s_1[i],\allowbreak s_2[i],...,s_k [i],s_{k+1}[i-1],s_{k+2}[i-1],\allowbreak...,s_M [i-1]\}$ and call them the {\it neighbor symbols} of $y_k[i]$. A sample $y_k[i]$ is then fully determined by the values of its neighbor symbols.
Note that the number of non-zero symbols in $\mathcal{V}(y_k[i])$ is the number of non-zero elements in the corresponding row of $\bm{D}$, giving
\begin{eqnarray}\label{eq:neighbors}
\left|\mathcal{V}(y_k[i])\right| =
\begin{cases}
k, & \text{when}~i=1; \\
M, & \text{when}~1\leq i\leq L; \\
M-k, &  \text{when}~i=L+1.
\end{cases}
\end{eqnarray}

\begin{figure*}[t]
  \centering
  \includegraphics[width=1.8\columnwidth]{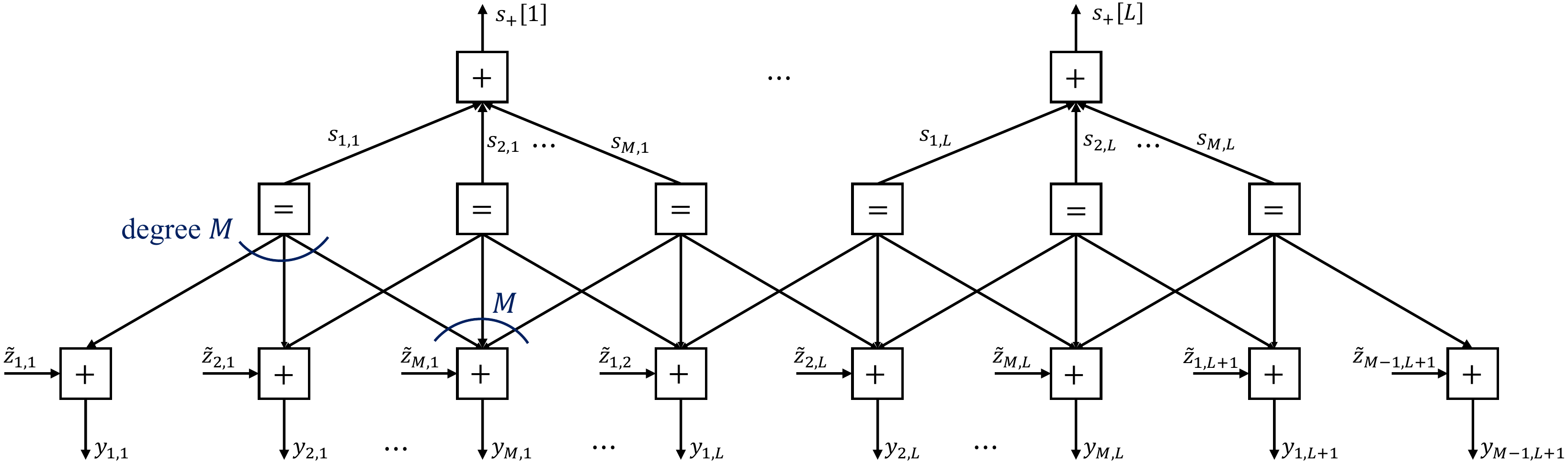}\\
  \caption{A graphical interpretation of the factorization in \eqref{eq:factorization}. To simplify notations, we denote $y_k[i]$, $z_k[i]$ and $s_m[i]$ by $y_{k,i}$, $z_{k,i}$ and $s_{m,i}$ in the figure, respectively.}
\label{fig:4}
\end{figure*}

Based on the factorization in \eqref{eq:factorization}, $f(\bm{y}|\bm{s})$ can be depicted by a graphical model \cite{kschischang2001factor,loeliger2007factor,PNC}. As shown in Fig.~\ref{fig:4}, we use a Forney-style factor graph \cite{loeliger2007factor} to represent the factorization. In particular, each edge in the graph corresponds to a variable in \eqref{eq:factorization}, i.e., a complex symbol $s_m[\ell]$, an observation $y_k[i]$, or a noise term $\tilde{z}_k[i]$. To simplify notations, we denote them by $s_{m,i}$, $y_{k,i}$, and $\tilde{z}_{k,i}$ in Fig.~\ref{fig:4}, respectively.

As can be seen, each sample $y_k[i]$ is related to a set of symbols $\mathcal{V}(y_k[i])$; each complex symbol $s_k[i]$, on the other hand, is related to $M$ samples. Thus, we duplicate each symbol $s_{m,i}$ for $M$ times and connect them to $M$ consecutive samples -- the equality constraint function ``$=$'' means that the values of the variables connecting to this function must be equal. The output degree of each equality constraint function is $M$, so is the input degree of each plus function ``$+$'' (except for the $M - 1$ samples at both ends of the packet). The target symbols $\bm{s}_+$ are shown at the top of Fig.~\ref{fig:4}.

The marginal likelihood function $f(\bm{y}|\bm{s[i]})$ can be obtained from the global likelihood function $f(\bm{y}|\bm{s})$ by a marginalization process operated on the factor graph, which can be implemented efficiently via the sum-product (SP) algorithm.
However, note in Fig.~\ref{fig:4} that we have a loopy graph. Moreover, the girth of the graph (the length of the shortest loop) is $4$. Such short loops prevent the sum-product algorithm from converging \cite{murphy2013loopy}. Even if they converge, the performance of the sum-product algorithm often degrades greatly, and the equilibrium posterior distribution is only an approximation of the true posterior distribution of the variables. Said in another way, the independence assumption of the extrinsic information passed along the edges no longer holds because the messages can circulate indefinitely around the loops.

\begin{figure*}[t]
  \centering
  \includegraphics[width=1.8\columnwidth]{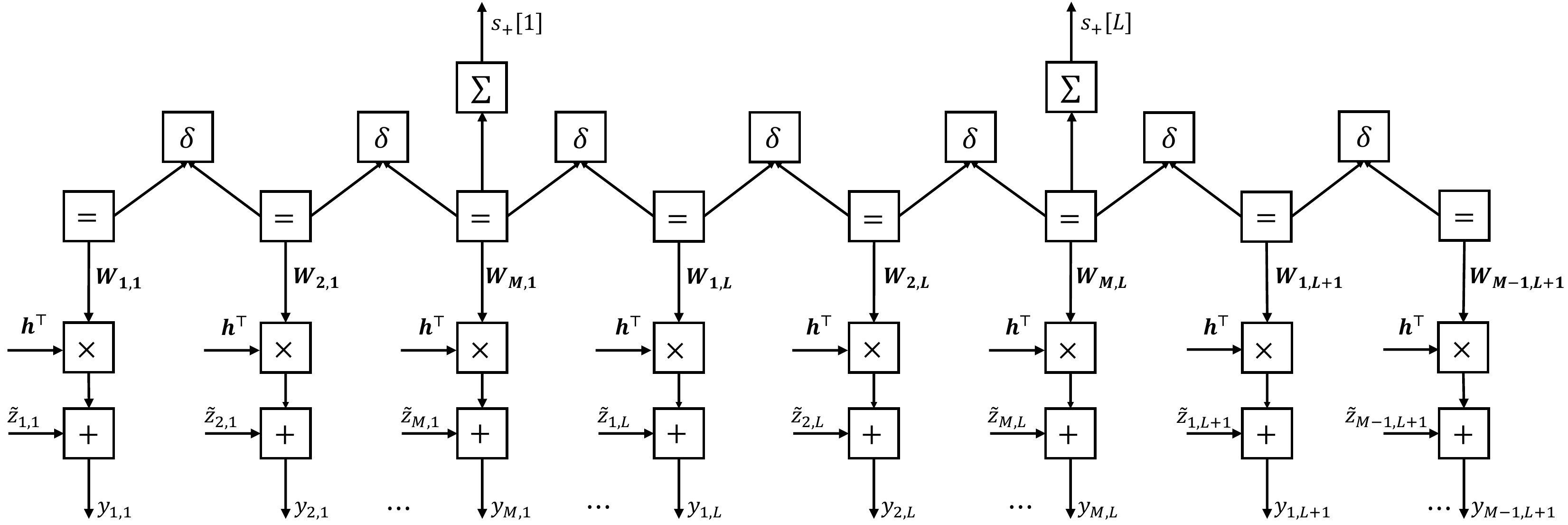}\\
  \caption{An equivalent tree structure to the loopy graph in Fig.~\ref{fig:4}. The variables connected to the same sample are clustered as a new high-dimensional variable. Each observation node is connected to a single variable node after clustering.}
\label{fig:5}
\end{figure*}

To circumvent this problem, below we transform the loopy graph in Fig.~\ref{fig:4} to a loop-free graph, at the expense of increasing the dimension of the variables \cite{kschischang2001factor}.
For each sample $y_k[i]$, we cluster all its neighbor symbols (i.e., the edges/variables that connect to $y_k[i]$ in Fig.~\ref{fig:4}) and construct a new higher-dimensional variable. Let us denote the new high-dimensional variable by $\bm{W}_{k,i}=\mathcal{V}(y_k[i])$. As illustrated in Fig.~\ref{fig:5}, each sample $y_k[i]$ is now connected to a single high-dimensional variable $\bm{W}_{k,i}$ after clustering and the loops are removed.

After clustering, the new high-dimensional variables $\bm{W}_{k,i}$ are correlated with each other because they contain common symbols. For example, $\bm{W}_{1,1}=\{s_1[1]\}$, $\bm{W}_{2,1}=\allowbreak \{s_1[1],\allowbreak s_2[1]\}$, and $\bm{W}_{3,1}\allowbreak =\{\allowbreak s_1[1],\allowbreak s_2[1],\allowbreak s_3[1]\}$, the common symbol between $\bm{W}_{1,1}$ and $\bm{W}_{2,1}$ is $s_1[1]$, that between $\bm{W}_{1,1}$ and $\bm{W}_{3,1}$ is $s_1[1]$, and that between $\bm{W}_{2,1}$ and $\bm{W}_{3,1}$ are $s_1[1]$, $s_2[1]$. Therefore, we have to add constraints among the new high-dimensional variables to ensure that the values of the common symbols are consistent across different variables.

In Fig.~\ref{fig:5}, the compatibility function $\delta$ is added on the adjacent variables to represent the above constraints. Specifically, for the two adjacent variables $\bm{W}$ and $\bm{W}^\prime$ connected to the same delta function, the compatibility function $\delta(\bm{W},\bm{W}^\prime)$ is defined as
\begin{eqnarray*}
\delta(\bm{W},\bm{W}^\prime)=\begin{cases}
1, & \hspace{-0.2cm}\text{if the values of all common symbols}\\
&\hspace{-0.2cm}\text{between}~\bm{W}~\text{and}~\bm{W}^\prime~\text{are equal}; \\
0, & \hspace{-0.2cm}\text{otherwise.}
\end{cases}
\end{eqnarray*}
That is, function $\delta$ is an on-off function that ensures that the messages passed from $\bm{W}$ to $\bm{W}^\prime$ and from $\bm{W}^\prime$ to $\bm{W}$ satisfy the constraint that the values of the common symbols between $\bm{W}$ and $\bm{W}^\prime$ are equal.

\begin{rem}
It is worth noting that adding compatibility functions between two adjacent variables is enough to depict all the constraints. For example, let us consider $\bm{W}_{1,1}$, $\bm{W}_{2,1}$ and $\bm{W}_{3,1}$. In Fig.~\ref{fig:5}, we only add compatibility functions $\delta(\bm{W}_{1,1},\bm{W}_{2,1})$ and $\delta(\bm{W}_{2,1},\bm{W}_{3,1})$. There is no need to add an extra compatibility function $\delta(\bm{W}_{1,1},\bm{W}_{3,1})$ between $\bm{W}_{1,1}$, $\bm{W}_{3,1}$ although they do have a common symbol $s_1[1]$. This is because $\bm{W}_{1,1}$ and $\bm{W}_{3,1}$ are independent conditioned on the compatibility functions $\delta(\bm{W}_{1,1},\bm{W}_{2,1})$ and $\delta(\bm{W}_{2,1},\bm{W}_{3,1})$. More specifically, $\delta(\bm{W}_{1,1},\bm{W}_{2,1})$ has ensured that the value of $s_1[1]$ in $\bm{W}_{1,1}$ equals that in $\bm{W}_{2,1}$ and $\delta(\bm{W}_{2,1},\bm{W}_{3,1})$ has ensured that the values of $s_1[1]$ in $\bm{W}_{2,1}$ equals that in $\bm{W}_{3,1}$. Thus, the values of $s_1[1]$ in $\bm{W}_{1,1}$ and $\bm{W}_{3,1}$ must be equal.
\end{rem}

Overall, the factor graph in Fig.~\ref{fig:5} presents a tree structure. Compared with Fig.~\ref{fig:4}, although the dimensions of the variables are $M$ times larger, the marginal likelihood function $f(\bm{y}|\bm{s[i]})$  can now be computed exactly via the sum-product algorithm thanks to the tree structure.

\subsection{Analog Message Passing and the SP-ML Estimator}\label{sec:IVC}
In standard sum-product algorithms, the messages passed on the edges are the probability mass function (PMF) of the variables associated with the edges, i.e., a probability vector of finite length \cite{kschischang2001factor,PNC}. This is because the transmitted symbols in digital communications are chosen from finite-size constellations. However, in the case of OAC, the transmitted symbols $\bm{s_m}$ are continuous complex numbers. Hence, the messages passed on the edges should be the probability density functions (PDFs) of the associated variables, which is a continuous function rather than a finite-length vector.

To enable message passing, a straightforward idea is to quantize the PDFs so that we can employ the digital message passing. The output of the SP algorithm is then the marginal PDF in a quantized form. However, quantization suffers from the ``curse of dimensionality'' -- when the dimension of the variables increases, the volume of the space increases exponentially fast. In order to get a sound result, a significantly larger number of quantization levels is required in each dimension compared with  the low-dimensional case \cite{noorshams2013belief}. In our problem, we have deliberately increased the dimensionality of the variables to remove the loops in Fig.~\ref{fig:4}. Thus, the standard sum-product algorithm with quantization cannot be used considering its prohibitive complexity.

Now that quantization is not an option, the problem is how to pass continuous PDFs along the edges of Fig.~\ref{fig:5}. A natural idea is then to parameterize the PDFs and pass their parameters \cite{loeliger2007factor,wang2019gaussian}.
In particular, we point out an important results from our companion paper \cite{BayesianOAC}: {\it if all inputs to a graph are Gaussian, then all the variables on the graph are multivariate Gaussian random variables.}
In Fig.~\ref{fig:5}, the only input to the factor graph is the likelihood function $f(y_{k,i}|\bm{W}_{k,i})$. We next show that this message is a multivariate Gaussian distribution with respect to $\bm{W_{k,i}}$.

Let us consider a complex random variable as a pair of real random variables: the real part and the imaginary part. Then, $\bm{W_{k,i}}$ can be viewed as a $2M$-dimensional real random variable and denoted by
$\bm{w_{k,i}} = \left(b^\mathfrak{r}_1,b^\mathfrak{r}_2,...,b^\mathfrak{r}_M,
b^\mathfrak{i}_1,b^\mathfrak{i}_2,...,b^\mathfrak{i}_M  \right)$.

As per \eqref{eq:samples}, we have
\begin{eqnarray*}
y_{k,i}\hspace{-0.2cm}&=&\hspace{-0.2cm}\sum_{k=1}^M (h^\mathfrak{r}_k+jh^\mathfrak{i}_k)(b^\mathfrak{r}_k+jb^\mathfrak{i}_k) + z^\mathfrak{r}_{k,i}  + jz^\mathfrak{i}_{k,i} \\
\hspace{-0.2cm}&=&\hspace{-0.2cm}\sum_{k=1}^M [(h^\mathfrak{r}_kb^\mathfrak{r}_k-h^\mathfrak{i}_kb^\mathfrak{i}_k+z^\mathfrak{r}_{k,i}) + j(h^\mathfrak{r}_kb^\mathfrak{i}_k+h^\mathfrak{i}_kb^\mathfrak{r}_k+z^\mathfrak{i}_{k,i})],
\end{eqnarray*}
where $z^\mathfrak{r}_{k,i}$, $z^\mathfrak{i}_{k,i}$ $\sim\mathcal{N}(0,\frac{N_0}{2d_k})$.
Thus, the likelihood function
\begin{eqnarray*}
&&\hspace{-0.7cm} f(y_{k,i}|\bm{w_{k,i}})\propto \exp\left\{-\frac{d_k}{N_0}[y^\mathfrak{r}_{k,i}-\sum_k (h^\mathfrak{r}_kb^\mathfrak{r}_k-h^\mathfrak{i}_kb^\mathfrak{i}_k)]^2 \right\} \\
&&\hspace{1.5cm} \times \exp\left\{-\frac{d_k}{N_0}[y^\mathfrak{i}_{k,i}-\sum_k (h^\mathfrak{r}_kb^\mathfrak{i}_k+h^\mathfrak{i}_kb^\mathfrak{r}_k)]^2 \right\} \\
&&\hspace{-0.7cm} \propto  \mathcal{N}\left(\frac{2d_k}{N_0}
\begin{bmatrix}
\bm{\beta_1} \\
\bm{\beta_2}
\end{bmatrix}
\begin{bmatrix}
y^\mathfrak{r}_{k,i} \\
y^\mathfrak{i}_{k,i}
\end{bmatrix},
\frac{2d_k}{N_0}
\begin{bmatrix}
\bm{\beta_1}\bm{\beta_1}^\top & \bm{\beta_1}\bm{\beta_2}^\top \\
\bm{\beta_2}\bm{\beta_1}^\top & \bm{\beta_1}\bm{\beta_1}^\top
\end{bmatrix}\right),
\end{eqnarray*}
where $\bm{\beta_1}$ and $\bm{\beta_1}$ are composed of channel coefficients, giving
\begin{eqnarray*}
\bm{\beta_1}=
\begin{bmatrix}
h^\mathfrak{r}_1 & h^\mathfrak{i}_1 \\
h^\mathfrak{r}_2 & h^\mathfrak{i}_2 \\
... & ... \\
h^\mathfrak{r}_M & h^\mathfrak{i}_M \\
\end{bmatrix},~~
\bm{\beta_2}=
\begin{bmatrix}
-h^\mathfrak{i}_1 & h^\mathfrak{r}_1 \\
-h^\mathfrak{i}_2 & h^\mathfrak{r}_2 \\
... & ... \\
-h^\mathfrak{i}_M & h^\mathfrak{r}_M \\
\end{bmatrix}.
\end{eqnarray*}

Since the likelihood functions $f(y_{k,i}|\bm{W}_{k,i})$, $\forall k,i$ are Gaussian, all the messages passed on Fig. \ref{fig:5} are multivariate Gaussian distributions. The Gaussian PDF can then be parameterized by its mean vector and covariance matrix -- passing these parameters is equivalent to passing the continuous PDF on the edges.

To ease reading, we summarize the main results of analog message passing in the following. The detailed proof can be found in Appendix B of our companion paper \cite{BayesianOAC}.
\begin{enumerate}[leftmargin=0.5cm]
\item The marginal likelihood function $f(\bm{y}|\bm{s[i]})$ is a multivariate complex Gaussian distribution of dimension $M$. The likelihood function $f(\bm{y}|{s}_+[i])$ is a single-variate complex Gaussian distribution.
\item A complex random variable can be viewed as a pair of real random variables (i.e., the real part and the imaginary part of the complex random variable). Thus, we can denote the $M$-dimensional complex Gaussian $f(\bm{y}|\bm{s[i]})$ by a $2M$-dimensional real Gaussian
\begin{eqnarray}\label{eq:Marginal}
\hspace{-0.7cm}&& f(\bm{y}|\bm{s[i]})\sim \\
\hspace{-0.7cm}&& \mathcal{N}\left(\bm{s[i]},
\bm{\mu_{s[i]}}=
\begin{bmatrix}
\bm{\mu^\mathfrak{r}_{s[i]}} \\
\bm{\mu^\mathfrak{i}_{s[i]}}
\end{bmatrix},
\bm{\Sigma_{s[i]}} =
\begin{bmatrix}
\bm{\Sigma^{\mathfrak{rr}}_{s[i]}} & \bm{\Sigma^{\mathfrak{ri}}_{s[i]}} \nonumber\\
\bm{\Sigma^{\mathfrak{ir}}_{s[i]}} & \bm{\Sigma^{\mathfrak{ii}}_{s[i]}}
\end{bmatrix}
\right),
\end{eqnarray}
where $\bm{\mu_{s[i]}}$  is a $2M$ by $1$ real vector consisting of the real and imaginary parts of the mean of $\bm{s[i]}$. That is, $\bm{\mu^\mathfrak{r}_{s[i]}}$ and $\bm{\mu^\mathfrak{i}_{s[i]}}$ are the real and imaginary parts of sequence $\mathbb{E}[\bm{s[i]}]^\top$. The covariance matrix $\bm{\Sigma_{s[i]}}$ is a $2M$ by $2M$ covariance matrix. The moment parameters $(\bm{\mu_{s[i]}},\bm{\Sigma_{s[i]}})$ can be computed by analog message passing.
\item Given the marginal likelihood function $f(\bm{y}|\bm{s[i]})$ in \eqref{eq:Marginal}, $f(\bm{y}|{s}_+[i])$ can be constructed by
\begin{eqnarray}\label{eq:splus}
\hspace{-0.7cm}&& f(\bm{y}|{s}_+[i])\sim \\
\hspace{-0.7cm}&& \mathcal{N}\left(
{s}_+[i],
\bm{\mu}_{s_+[i]}\!=\!
\begin{bmatrix}
\mu^\mathfrak{r}_{s_+[i]} \\
\mu^\mathfrak{i}_{s_+[i]}
\end{bmatrix},
\bm{\Sigma}_{s_+[i]} \!=\!
\begin{bmatrix}
\Sigma^{\mathfrak{rr}}_{s_+[i]} & \Sigma^{\mathfrak{ri}}_{s_+[i]} \nonumber\\
\Sigma^{\mathfrak{ir}}_{s_+[i]} & \Sigma^{\mathfrak{ii}}_{s_+[i]}
\end{bmatrix}
\right),
\end{eqnarray}
where
\begin{eqnarray*}
\hspace{-0.7cm}&&\mu^\mathfrak{r}_{s_+[i]} = \bm{1}^\top\bm{\mu^\mathfrak{r}_{s[i]}},~~ \mu^\mathfrak{i}_{s_+[i]} = \bm{1}^\top\bm{\mu^\mathfrak{i}_{s[i]}}, \\
\hspace{-0.7cm}&&\Sigma^{\mathfrak{rr}}_{s_+[i]} = \bm{1}^\top  \bm{\Sigma^{\mathfrak{rr}}_{s[i]}}  \bm{1},~~
\Sigma^{\mathfrak{ri}}_{s_+[i]} = \bm{1}^\top  \bm{\Sigma^{\mathfrak{ri}}_{s[i]}}  \bm{1}, \\
\hspace{-0.7cm}&&\Sigma^{\mathfrak{ir}}_{s_+[i]} = \bm{1}^\top  \bm{\Sigma^{\mathfrak{ir}}_{s[i]}}  \bm{1}, ~~
\Sigma^{\mathfrak{ii}}_{s_+[i]} = \bm{1}^\top  \bm{\Sigma^{\mathfrak{ii}}_{s[i]}}  \bm{1}.
\end{eqnarray*}
\item Following \eqref{eq:splus}, an SP-ML estimator can be designed, as given in Definition~\ref{defi:SP_ML}.
\end{enumerate}

\begin{defi}[SP-ML estimation]\label{defi:SP_ML}
The WMFS scheme gives us the whitened samples $\bm{y}$ in \eqref{eq:samplesMat}. To estimate the designed sequence $\bm{s_+}$, an SP-ML estimator first computes the moment parameters of the multivariate Gaussian $f(\bm{y}|\bm{s[i]})$, $\forall i$, by an analog sum-product process, and then estimates each element of $\bm{s_+}$ by
\begin{eqnarray}\label{eq:SPML}
\hat{s}^\text{ml}_+[i] = \bm{1}^\top\bm{\mu^\mathfrak{r}_{s[i]}} +j \bm{1}^\top\bm{\mu^\mathfrak{i}_{s[i]}},
\end{eqnarray}
where $\bm{\mu_{s[i]}}= [\bm{\mu^\mathfrak{r}_{s[i]}},\allowbreak\bm{\mu^\mathfrak{i}_{s[i]}}]^\top$ is the mean of $f(\bm{y}|\bm{s[i]})$.
\end{defi}

The reason behind \eqref{eq:SPML} is as follows. It has been shown that $f(\bm{y}|{s}_+[i])$ is conditionally Gaussian. As per the ML rule, we should choose the mean of $f(\bm{y}|{s}_+[i])$ as the estimate of $s_+[i]$ as it maximizes the likelihood function. This gives us \eqref{eq:SPML}.

\begin{rem}[maximum likelihood sequence estimation (MLSE) versus Bahl-Cocke-Jelinek-Raviv (BCJR)]
The ML estimators in \eqref{eq:ML1} and \eqref{eq:ML2} aim to find the ML sequence $\bm{s_+}$ in the space $\mathbb{C}^L$. In the language of digital communications, they are ML optimal in the sense of MLSE \cite{MLSE}. On the other hand, the ML estimator in \eqref{eq:SPML} aims to maximize the likelihood function of each element of $\bm{s_+}$. Thus, it is ML optimal in the BCJR \cite{BCJR} sense.

When we perform ML estimation in digital communications, MLSE-optimal and BCJR-op\-timal are different criteria because the former minimizes the block error rate (BLER) while the latter minimizes the bit error rate (BER). For the ML estimation in OAC, however, the two criteria are equivalent.
The reason for this discrepancy is again that the discrete constellations used in digital communications serve as a kind of prior information to the receiver while an OAC receiver has no prior information at all.

More specifically, let us consider the message passing in Fig.~\ref{fig:5}.
For the ML estimation in OAC, we have shown that all the messages passed on the graph, including $f(\bm{y}|\bm{s_+})$ and $f(\bm{y}|s_+[i])$, are Gaussian. Thus, the ML sequence $\bm{s_+}$ also gives us the ML $s_+[i]$, $\forall i$, after marginalization. As a result, MLSE-optimal and BCJR-op\-timal are equivalent, and the ML estimators in \eqref{eq:ML1}, \eqref{eq:ML2}, and \eqref{eq:SPML} are identical.
In contrast, when we perform ML estimation in digital communications, the prior information imposes each transmitted symbol to belong to a finite constellation. As a consequence, the messages passed on the graph are Gaussian mixtures, and neither $f(\bm{y}|\bm{s_+})$ nor $f(\bm{y}|s_+[i])$ is Gaussian distributed. As a result, MLSE-optimal and BCJR-op\-timal are different criteria.
\end{rem}

\textbf{Computational complexity} -- Finally, we evaluate the computational complexity of the SP-ML estimator.
With analog message passing, messages passed on the graph are simply the parameters of the Gaussian distributions instead of continuous Gaussian PDFs.
Computations involved in analog message passing are simply 1) the sum of 2M-dimensional vectors/matrices, and 2) 2M-dimensional matrix inversion.
Therefore, the computational complexity of the SP-ML estimator is $\Omega(LM^2\log M)$.
If we fix $M$ as a constant, the SP-ML estimator significantly reduces the computational complexity of ML estimation from $\Omega(L^2\log L)$ to $\Omega(L)$.

\subsection{Aligned-Sample Estimator}
As stated in the beginning of this section, another benefit of the WMFS scheme is that it yields a sequence of samples wherein the indexes of symbols from different devices are consistent.
Specifically, let us consider the outputs of the $M$-th matched filter.

Let $k = M$ in \eqref{eq:samples}, we have $\mathcal{V}(y_M[i])\allowbreak=\{s_1[i],\allowbreak s_2[i],...,\allowbreak s_M[i]\}$, and
\begin{eqnarray}\label{eq:samplesMth}
y_M[i]=\sum_{m=1}^M h^\prime_{M}[i]s_m[i]+z_M[i],
\end{eqnarray}
where $z_M[i]\sim\mathcal{CN}(0,N_0/d_M)$ and $d_M$ is the duration of the $M$-th matched filter. As can be seen, unlike the outputs of other matched filters, the neighbor symbols of $y_M[i]$ have the same index (that is, the symbol indexes are aligned within the integral interval of the $M$-th matched filter). Therefore, we can utilize the outputs of the $M$-th matched filter to devise an aligned-sample estimator.

\begin{defi}[Aligned-sample estimator for misaligned OAC]\label{defi:11}
Given the outputs of the WMFS $\{y_k[i]\}$, aligned-sample estimator estimates the desired sequence $\bm{s}_+\in\mathcal{C}^L$ symbol-by-symbol by
\begin{eqnarray}\label{eq:alignedEstimator}
\hat{s}_+[i]=y_M[i].
\end{eqnarray}
\end{defi}

Eq. \eqref{eq:samplesMth} is an underdetermined equation since we have one equation for $M$ unknowns, and the estimator in \eqref{eq:alignedEstimator} is our best prediction about $\hat{s}_+[i]$. When there is no or mild channel-gain misalignment (i.e., $h^\prime_{M}\to 1$), the estimator \eqref{eq:alignedEstimator} is supposed to perform very well.

\section{Simulation Results}\label{sec:V}
This section evaluates the system performance of FEEL with misaligned OAC considering two estimators at the receiver: the ML estimator and the aligned-sample estimator.
In particular, for ML estimation, we use our SP-ML estimator since the ML estimators in \eqref{eq:ML1} and \eqref{eq:ML2} are computationally prohibitive.\footnote{We have performed additional simulations to validate that the three estimators in \eqref{eq:ML1}, \eqref{eq:ML2}, and \eqref{eq:SPML} are equivalent (using a much shorter packet length $L$). The simulation results are omitted here to conserve space.}
We implement a FEEL system wherein $40$ devices collaboratively train a convolution neural network (CNN) to solve the CIFAR-10 classification task \cite{CIFAR10}.
The CIFAR-10 dataset has a training set of $50,000$ examples and a test set of $10,000$ examples in 10 classes. Each example is a $32\times 32$ colour image.
The non-i.i.d. training examples are assigned to the $40$ devices in the following manner:
1) we first let each device randomly sample $1,000$ samples from the dataset;
2) for the remaining $10,000$ examples, we sort them by their labels and group into $40$ shards of size $250$ \cite{FL2}. Each device is then assigned one shard.

The implemented CNN is a ShuffleNet V2 network \cite{Shufflenet} with $d=1.26\times 10^6$ parameters (this corresponds to $6.32\times 10^5$ complex values).
In each iteration, we assume $M = 4$ devices are active and participate in the training. Each device will train the global model locally for $5$ epochs and then transmit the model-update to the PS in $44$ packets (the packet length is $L=1.44\times 10^4$) in each iteration. The packets from different transmitters overlap at the PS with time and channel-gain misalignments, and the PS employs the ML and aligned-sample estimators to estimate the arithmetic sum of transmitted symbols, i.e., $\bm{s_+}$ (and hence, $\bm{\theta_+}$). The estimated arithmetic sum $\bm{\theta_+}$ is then used to update the global model, as per \eqref{eq:II2}.
All the source codes are available online \cite{sourcecode}.

\begin{figure}[t]
  \centering
  \includegraphics[width=0.8\columnwidth]{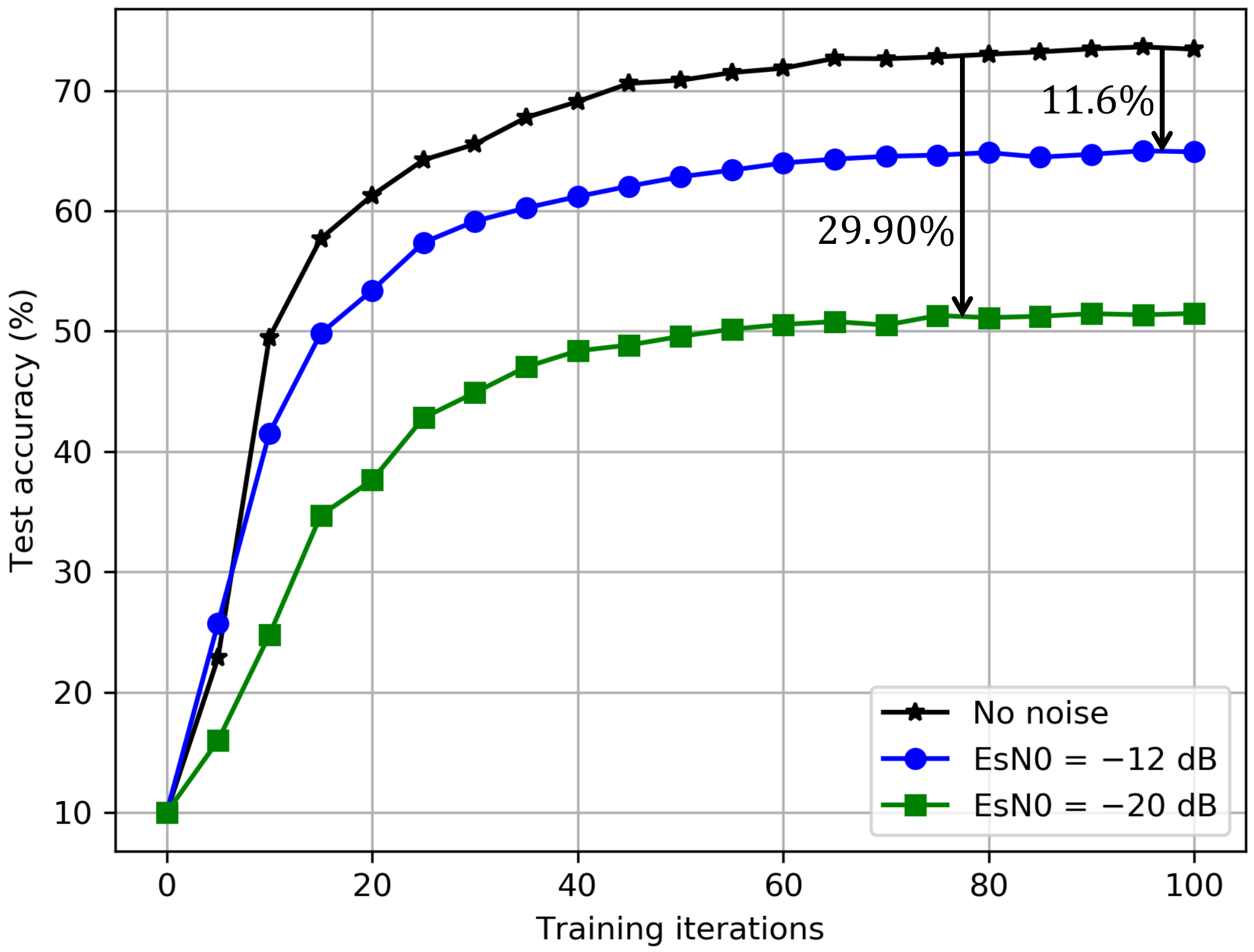}\\
  \caption{Test accuracy of the learned model over the course of training, with and without noise. There is no time or channel-gain misalignment. We use the ML estimator at the PS.}
\label{fig:sim1}
\end{figure}

The metric we use to assess the performance of an estimator is the {\it test accuracy}. Specifically, when operated with a given estimator, we will train the global model for $100$ iterations and take the prediction accuracy of the learned model on the test set as the performance indicator of the estimator.
An example is given in Fig.~\ref{fig:sim1}.

As shown in Fig.~\ref{fig:sim1}, we run the FEEL system for $100$ iterations with and without noise, and plot the test accuracy over the course of training. There are no misalignments in this simulation and ML estimator is used at the PS.
First, the dark curve corresponds to the noiseless case and the test accuracy after $100$ iterations is $73.43\%$.
We point out that this is the global optimal test accuracy since the MAC is ideal, i.e., there is no time misalignment, channel-gain misalignment, or noise in the MAC.
The other two curves in Fig.~\ref{fig:sim1} correspond to the learning performance when noise is presented in the received signal.
In particular, noise is added according to a given EsN0, i.e., the average received energy per symbol to noise power spectral density ratio, defined as
\begin{eqnarray}
\text{EsN0}=\frac{\mathbb{E}_i\left[\left|\sum_{m=1}^{M}h^\prime_m s_m[i] \right|^2\right]}{N_0}.
\end{eqnarray}
As shown, noise is detrimental to the test accuracy after convergence.
Compared with the noiseless case, the test accuracy drops by $11.6\%$ with an EsN0 of $-12$ dB, and by $29.9\%$ with an EsN0 of $-20$ dB.

In addition to noise, we next introduce time misalignment into the received signal.
The received signal is given in \eqref{eq:samples}, where the noise term is $z[i]\sim\mathcal{CN}(0,N_0/d_k)$, and we set the residual channel gain to $h^\prime_m=1$, $\forall~m$. Without loss of generality, symbol duration is set to $T = 1$.
As can be seen from \eqref{eq:samples}, time offsets $\{\tau_m:m=1,2,...,M\}$ determine the noise power of samples from different matched filters (since $d_k=\tau_{k+1}-\tau_k$).
In the simulation, the time offsets $\tau_m$, $\forall~m$, are set in the following manner: first, we fix a maximum time offset $\tau_M$ (and hence $d_M$); then, we generate the other time offsets uniformly in $(0,\tau_M)$.

\begin{figure}[t]
  \centering
  \includegraphics[width=0.8\columnwidth]{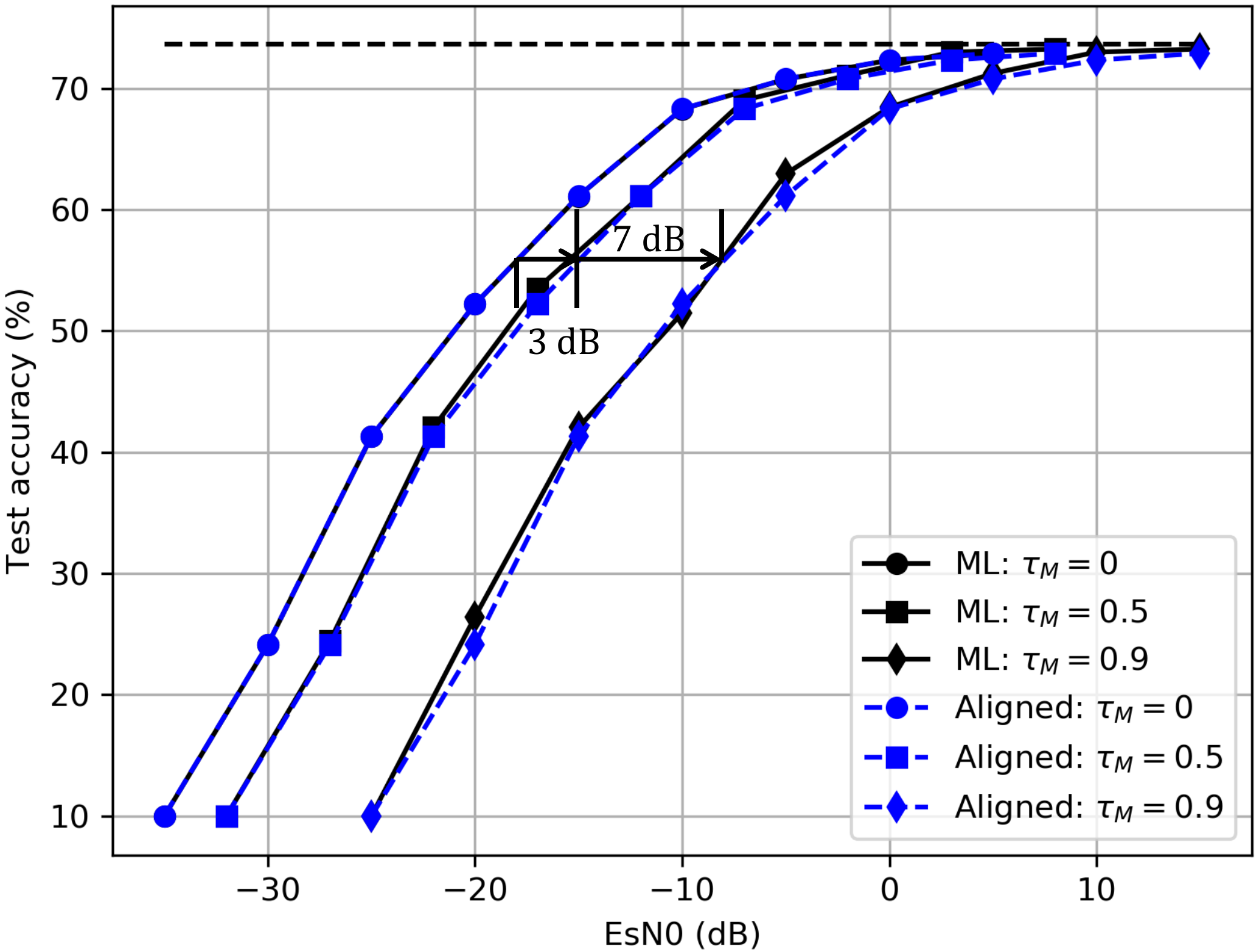}\\
  \caption{Test accuracy (after $100$ iterations) of the asynchronous OAC under various EsN0. There is no channel-gain misalignment in the simulation and we use both the ML estimator and the aligned-sample estimator. }
\label{fig:sim2}
\end{figure}

The simulation results are presented in Fig.~\ref{fig:sim2}. When $\tau_M=0$, there is neither time nor channel-gain misalignment in the received signal. The ML estimator and the aligned-sample estimator are equivalent in this case, and yield the same test accuracy. As we increase $\tau_M$, the performance of both estimators deteriorate.
\begin{enumerate}[leftmargin=0.5cm]
\item For the aligned-sample estimator, the performance deterioration is easy to understand since the inputs to the estimator are the outputs of the $M$-th matched filter $y_M[i]$. As a result, the performance of the aligned-sample estimator is governed by the maximum time offset $\tau_M$ -- the larger the $\tau_M$, the worse the performance.
As shown in Fig.~\ref{fig:sim2}, the introduction of time misalignment results in an EsN0 penalty for the aligned-sample estimator. When $\tau_M=0.5$, the EsN0 penalty is $3$ dB because $d_M$ is reduced by a factor of $2$ (from $1$ to $0.5$). Likewise, the EsN0 penalty is $10$ dB when $\tau_M=0.9$ since $d_M$ is reduced by $10$ times (from $1$ to $0.1$).
\item For different $\tau_M$, the performance gain of the ML estimator over the aligned-sample estimator is negligible.
The aligned-sample estimator utilizes only the outputs of the $M$-th matched filter.
The ML estimator, on the other hand, utilizes the outputs of all matched filters and attempts to estimate the mostly likely $\bm{s}_+$.
It turns out that both estimators yield nearly the same performance when there is only time misalignment and noise.
\end{enumerate}

It should be noted that the above result does not imply that the samples of the matched filters other than the $M$-th one are useless, because the ML estimator cannot achieve the same performance as the aligned-sample estimator using only the outputs of the $M$-th matched filter: recall from \eqref{eq:alignedEstimator} that the outputs of the $M$-th matched filter results in a set of underdetermined equations. If we perform ML estimation based on the samples in \eqref{eq:alignedEstimator}, the estimation error can be arbitrarily large as ${s}_+[i]$ can be any value.

\begin{rem}[Error propagation]
In the misaligned OAC, ML estimation boils down to MUE and faces an infinitely large estimation space.
It is then very susceptible to noise and suffers from error propagation.
Take the SP-ML estimator for instance. In the forward message passing, the successful estimation of a likelihood function (of a multivariate variable) hinges on the accurate estimations of the likelihood functions on the left.
When a sample is contaminated by noise, the mean of the likelihood function deviates from the true value of the noiseless sample. This estimation error will be propagated along the tree all the way to the rightmost leaf, because there are no known messages (i.e., prior information) in between to alleviate/correct the error.
This can be one cause of the results in Fig.~\ref{fig:sim2}.
\end{rem}

In the third simulation, let us further introduce channel-gain misalignment into the received signal \eqref{eq:samples}.
For each device, the residual channel gain is $h_m=|h_m|e^{j\phi_m}$.
We set $|h_m|=1$, $\forall m$, and focus on the impact of the phase offsets $\phi_m$ only. In particular, we assume $\{\phi_m: m=1,2,...,M\}$ are uniformly distributed in $(0,\phi)$, where $\phi$ is the maximum phase offset (i.e., $\phi_m\sim U(0,\phi)$). It is worth noting that $\phi_m$ can be any distribution in general.

\begin{figure}[t]
  \centering
  \includegraphics[width=0.8\columnwidth]{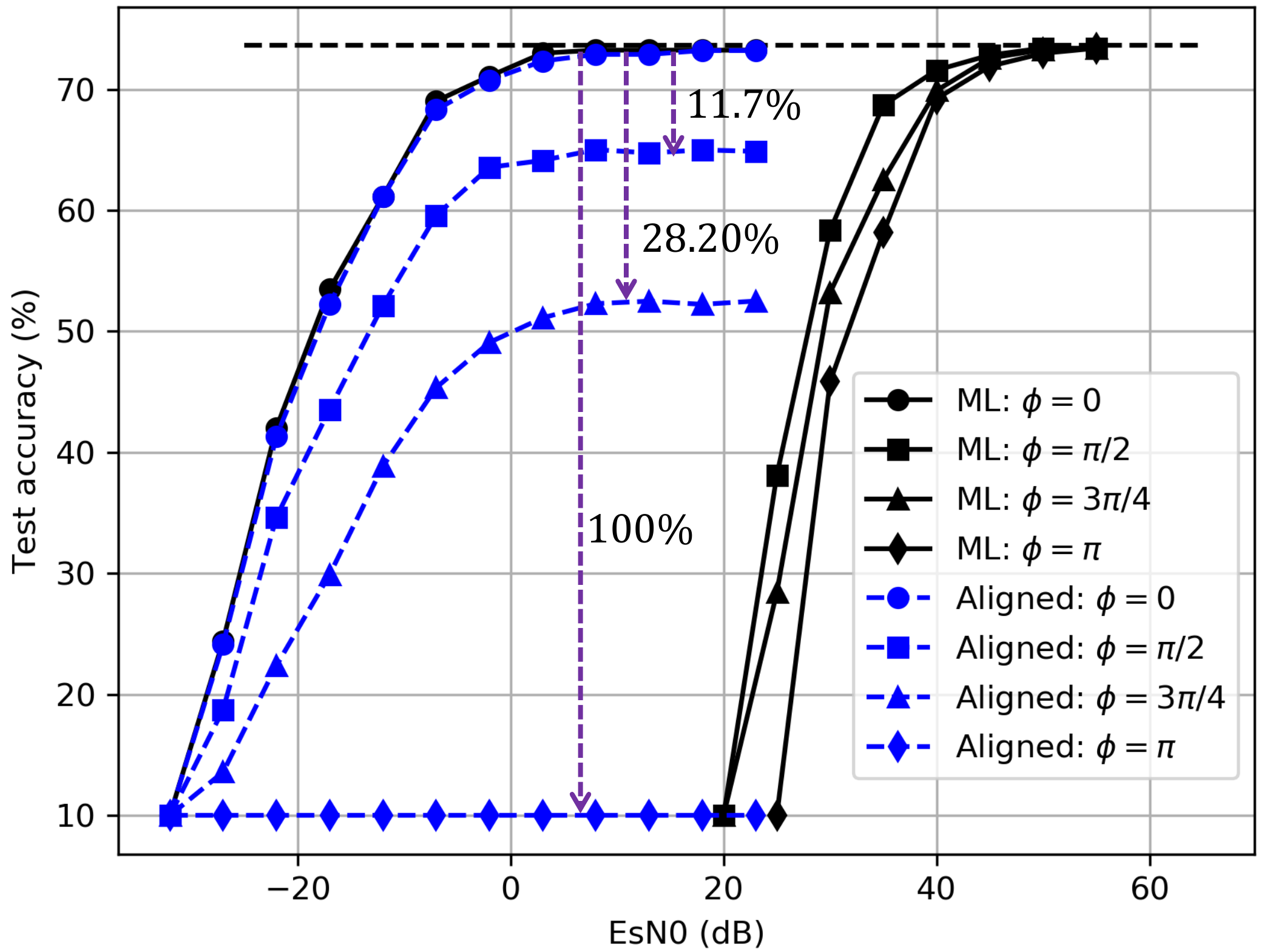}\\
  \caption{Test accuracies of the ML and aligned-sample estimators under different degrees of phase misalignments. The maximum time offset $\tau_M=0.5$ and the maximum phase misalignment $\phi=0$ (no), $\pi/2$ (mild), $3\pi/4$ (moderate), and $\pi$ (severe), respectively.}
\label{fig:sim3}
\end{figure}

Fig.~\ref{fig:sim3} presents the test accuracy of the ML and aligned-sample estimators versus EsN0 (in dB), wherein the maximum time offset $\tau_M$ is fixed to $0.5$ and the maximum phase offset $\phi=0$ (no phase misalignment), $\pi/2$ (mild), $3\pi/4$ (moderate), and $\pi$ (severe), respectively.

\begin{rem}
A caveat here is that $\phi$ is the maximum phase offset -- the phase offsets of all devices are uniformly distributed in $[0,\phi]$. If we look at the phase misalignment between any two devices, however, the average pair\-wise-pha\-se-mis\-alignm\-ent is only $\phi/3$. That is why we classify $\pi/2$ as mild phase misalignment because the average pa\-irwi\-se-ph\-ase-m\-isali\-gn\-ment is only $\pi/6$.
\end{rem}

We have the following observations from Fig.~\ref{fig:sim3}:
\begin{enumerate}[leftmargin=0.5cm]
\item When there is no phase misalignment ($\phi=0$), the two curves coincide just as in Fig.~\ref{fig:sim2}.
\item When there is mild phase misalignment ($\phi=\pi/2$), the aligned-sample estimator suffers from two penalties: i) a small EsN0 penalty for about $5$ dB, i.e., we need a $5$ dB higher EsN0 to achieve the same test accuracy; ii) a $11.7\%$ test-accuracy loss, i.e., the test accuracy after convergence is $11.7\%$ less than the phase-aligned case.

The ML estimator, on the other hand, suffers from a large EsN0 penalty. The reason behind is that the phase misalignment enhances the error/noise propagation in ML estimation, which we refer to as {\it noise enhancement}. As a result, ML estimation does not work in the low-EsN0 regime when there is phase misalignment.
On the bright side, ML estimator performs very well in the high-EsN0 regime: it suffers from no test-accuracy loss -- the test accuracy after convergence is the same as the phase-aligned case.
\item When we further increase the maximum phase misalignment $\phi$, the aligned-sample estimator suffers from larger EsN0 and test-accuracy penalties. In the case of moderate phase misalignment ($\phi=3\pi/4$), the test-accuracy penalty is up to $28.2\%$. In the case of severe phase misalignment ($\phi=\pi$), the learning diverges with the aligned-sample estimator.

In contrast, the ML estimator is more robust to moderate and severe phase misalignments in the high-EsN0 regime -- there is no test-accuracy loss and only a small EsN0 penalty.
\end{enumerate}

\begin{figure}[t]
  \centering
  \includegraphics[width=0.8\columnwidth]{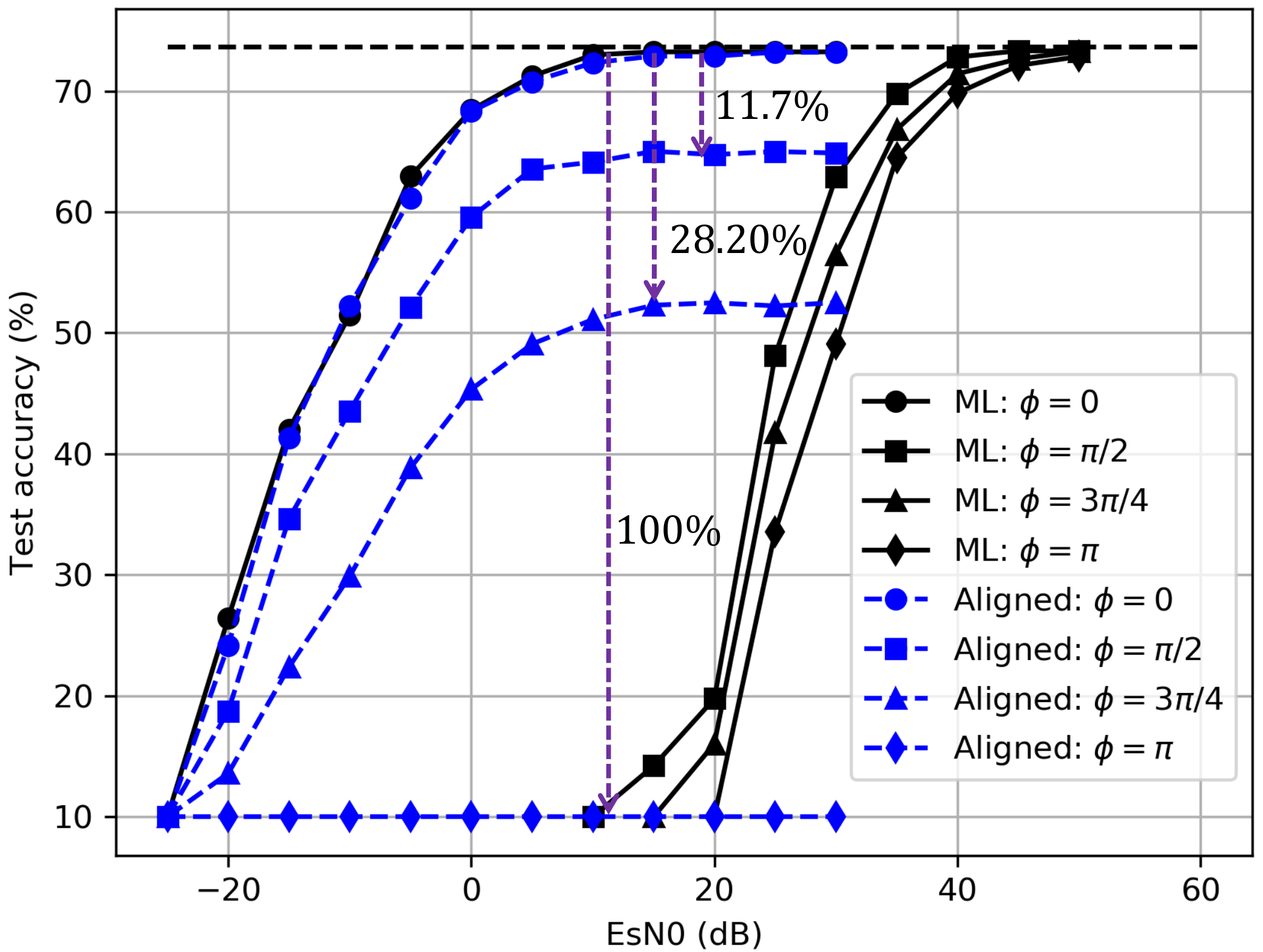}\\
  \caption{Test accuracies of the ML and aligned-sample estimators under different degrees of phase misalignments. The maximum time offset $\tau=0.9$ and the maximum phase misalignment $\phi=0$ (no), $\pi/2$ (mild), $3\pi/4$ (moderate), and $\pi$ (severe), respectively.}
\label{fig:sim4}
\end{figure}

Fig.~\ref{fig:sim3} studies the impact of phase misalignment under a fixed maximum time offset $\tau_M=0.5$. Next, we consider a larger time offset $\tau_M=0.9$ and repeat the simulations in Fig.~\ref{fig:sim3}. The simulation results are presented in Fig.~\ref{fig:sim4}.

For the aligned-sample estimator, time offset only incurs an EsN0 penalty. Thus, each performance curve of the aligned-sample estimator in Fig.~\ref{fig:sim4} is simply a right-shift of the corresponding curve in Fig.~\ref{fig:sim3} by $7$ dB.

On the other hand, we observe that ML estimation benefits from larger time misalignment when there is phase misalignment.
For example, with a mild phase misalignment ($\phi=\pi/2$), the EsN0 gain is about $2$ dB when the maximum time offset $\tau_M$ is increased from $0.5$ to $0.9$, as shown in Fig.~\ref{fig:sim3} and Fig.~\ref{fig:sim4}. In contrast, when there is no phase misalignment, ML estimation suffers from larger time misalignment, as shown in Fig.~\ref{fig:sim2}.

To conclude this section, we summarize the main simulation results as follows.
\begin{enumerate}[leftmargin=0.5cm]
\item When there is no phase misalignment, the ML and aligned-sample estimators are on equal footing as far as the learning performance is concerned.
\item When there is mild or moderate phase misalignment, the aligned-sample estimator outperforms the ML estimator in the low-EsN0 regime, but is worse than the ML estimator in the high-EsN0 regime.
\item When there is severe phase misalignment, the aligned estimator leads to divergence of learning, but ML estimation still works in the high-EsN0 regime.
\end{enumerate}

\section{Time-domain Realization versus Frequency-domain Realization}\label{sec:VI}
In this paper, we considered a time-domain realization of OAC.
OAC can also be realized in the frequency domain via OFDM.
An interesting direction to extend the current work is to compare the two realizations and their abilities to combat misalignments. On the other hand, our study in this paper also sheds light on the frequency-domain realization of OAC. A brief comparison between the two realizations is given below to provide some operational insights.

Time-domain or frequency-domain realization of wireless communication systems has been a long-standing debate.
When it concerns misalignments, the time-domain realization is sensitive to time offsets among edge devices, while the frequency-domain realization is sensitive to the carrier frequency offsets (CFOs) among edge devices.
\begin{enumerate}[leftmargin=0.5cm]
\item Time misalignment. With time-domain realization, time misalignment leads to an EsN0 penalty to the learning performance, as verified in Section \ref{sec:V}.
OFDM, on the other hand, deliberately introduces redundancies known as the cyclic prefix (CP) and transforms the time offset of each device $\tau_m$ to the frequency domain as extra phase offsets $e^{j2\pi \ell\tau_m/LT}$, $\ell=1,2,...,L$, on the $L$ subcarriers.
If we look at one subcarrier, it is equivalently a synchronous time-domain realization with phase misalignments -- the frequency domain samples are the same as \eqref{eq:samplesMth} by setting $h^\prime_m=e^{j2\pi \ell\tau_m/LT}$ and $d_M=T$. It is then an underdetermined equation and we can use the aligned-sample estimator in Definition \ref{defi:11} to estimate the arithmetic sum.
We emphasize that the phase misalignment can be severe for large $\tau_M$ (maximum time offset) and $\ell$ (subcarrier index). For example, when $\tau_M=T$ and $\ell=L$, the phase misalignment is up to $2\pi$.

When there is no CFO, the performance of the OFDM system for the misaligned OAC can be predicted by the performance of the aligned-sample estimator in Fig. \ref{fig:sim3} after left-shifting for $3$ dB (i.e., let $\tau_M=0$).

To summarize, OFDM systems introduce redundancies to transform time misalignment to phase misalignment. It is equivalent to trading off test-accuracy loss for EsN0 loss.
It is worth noting that we can also insert some redundancies, e.g., pilots, in the time-domain realizations to improve the performance of ML estimation
\item CFO. For the time-domain realization considered in this paper, residual CFO in the overlapped signal simply introduces additional channel-gain misalignments among devices.
For a frequency-domain realization, however, CFO introduces inter-carrier interference (ICI), a dual problem of the inter-symbol and inter-user interferences in the time-domain realization of OAC.
As a result, we have to devise an ML estimator to combat ICI, and ML estimation in OFDM with residual CFO falls into the same scope of ML estimation in time-domain realization with time misalignment.
In this context, the analysis in this paper about the properties of ML estimation for the misaligned OAC still holds, and the SP-ML estimator devised in Section \ref{sec:IV} can also be used in OFDM systems to perform ML estimation.
\end{enumerate}

\section{Conclusion}\label{sec:Conclusion}
As a joint computation-and-communication technique, OAC exploits the property of the MAC that its output is the arithmetic sum of the inputs. OAC is an efficient scheme to speed up the uplink aggregation of models from the edge devices in FEEL.
This paper filled the research gap of the misaligned OAC by devising two estimators, an SP-ML estimator and an aligned-sample estimator, to estimate the arithmetic sum of the symbols from different devices in the face of channel-gain and time misalignments.
The underpinning of the proposed estimators is an oversampled matched filtering and sampling scheme that yields:
a) whitened samples with alleviated inter-symbol and inter-user interferences;
b) a subsequence of samples wherein the indexes of transmitted symbols from different devices are aligned.

\textbf{The ML estimator} -- The whitened samples in a) allows us to construct a factor graph with a simple structure to represent the compositions of the samples, whereby an SP-ML estimator can be devised to compute the ML estimate of the arithmetic sum from an analog message passing process.
Compared with conventional ML estimator that is computationally prohibitive, the complexity of the SP-ML estimator grows linearly with the packet length, and hence, is computationally more efficient.

In the OAC system, symbols transmitted from the edge devices are continuous values instead of discrete constellations. Therefore, we have no prior information about the transmitted symbols and ML estimation is the only option.
In this context, the arithmetic-sum estimation boils down to MUE and the estimation space is infinitely large.
Two problems with the ML estimation are error propagation and noise enhancement. Specifically, the estimation error introduced by noise in a sample can propagate to other samples, causing larger and larger deviations from the true values for all the samples in between. The error propagation is further intensified by phase misalignment since the error/noise can be amplified in the propagation.

As a result, ML estimation does not perform well in the low EsN0 regime, especially when there is phase misalignment. To address this problem, a possible solution is to insert pilots in the transmitted symbols to cut off the error propagation.

\textbf{The aligned-sample estimator} -- The subsequence of the ``aligned'' samples in b), on the other hand, allows an aligned-sample estimator to be used in misaligned OAC.
The upside of the aligned-sample estimator is that it does not suffer from error propagation and noise enhancement issues; and hence, it emerges a good alternative to the ML estimator in the low-EsN0 regime.
The downsides, however, are that it suffers from both phase misalignment and time misalignment -- phase misalignment causes a test-accuracy loss and time misalignment causes an EsN0 penalty to the aligned-sample estimator.

The computational complexities of both ML and aligned-sample estimators grow linearly with the packet length. The aligned-sample estimator is preferred in the low-EsN0 regime and the ML estimator is preferred in the high-EsN0 regime.

\appendices
\section{}\label{sec:AppA}
In this appendix, we generalize the system model in Section \ref{sec:II} by considering fast fading channels $\tilde{h}_m(t)$ and residual CFO in the received signal $r(t)$.

At each transmitter, to compensate the fast channel fading and CFO between the transmitter and the receiver, $\alpha_m$ in \eqref{eq:II3} is set as $\alpha_m(t)= e^{-j\bar{\varepsilon}_m t}/\bar{h}_m(t)$, where $\bar{\varepsilon}_m$ is the estimated CFO at the $m$-th device.

The received signal $r(t)$ is then given by
\begin{eqnarray}\label{eq:App1}
r(t)=\sum_{m=1}^M \tilde{h}_m(t){e^{j\tilde{\varepsilon}_m t}} x_m(t-\tau_m) +z(t).
\end{eqnarray}
Unlike \eqref{eq:II4}, each $\tilde{h}_m(t)$ is now a fast fading channel and $\tilde{\varepsilon}_m$ is the CFO between the $m$-th device and the PS.

Substituting \eqref{eq:II3} into \eqref{eq:App1} gives us
\begin{eqnarray}\label{eq:Apprt}
r(t)\hspace{-0.3cm}&=&\hspace{-0.3cm}\sum_{m=1}^M\tilde{h}_m(t) e^{j\tilde{\varepsilon}_m t}\alpha_m(t)\sum_{\ell=1}^L s_m[\ell]p(t-\tau_m-\ell T) + z(t) \nonumber\\
\hspace{-0.3cm}&=&\hspace{-0.3cm} \sum_{\ell=1}^L \sum_{m=1}^M {h}'_m(t) e^{j{\varepsilon}'_m t} s_m[\ell]p(t-\tau_m-\ell T) + z(t),
\end{eqnarray}
where $h'_m(t)=\tilde{h}_m(t)/\bar{h}_m(t)$ is the residual channel-fading coefficient and ${\varepsilon}'_m= \tilde{\varepsilon}_m - \bar{\varepsilon}_m$ is the residual CFO between the $m$-th device and the PS.
As can be seen, CFO introduces additional channel-gain misalignments among devices.

With the revised signal $r(t)$, the discrete samples can be written in the same form as \eqref{eq:sample1} with different coefficients $c_{m,k}[i]$ and $c^\prime_{m,k}[i]$, giving,
\small
\begin{eqnarray}
\label{eq:A1}
\hspace{-0.65cm}&& c_{m,k}[i] = \frac{1}{T} \int_{(i-1)T+\tau_{k}}^{(i-\mathbbm{1}_{m>k})T+\tau_{m}} h'_m(t) e^{j\varepsilon^\prime_m\zeta}\,d\zeta = \frac{1}{j\varepsilon^\prime_m T}\times \\
\hspace{-0.65cm}&& \left(e^{j\varepsilon^\prime_m[(i\!-\!\mathbbm{1}_{m>k})T+\tau_{m}]}\!-\!e^{j\varepsilon^\prime_m[(i\!-\!1)T+\tau_{k}]}\right)
\int_{(i-1)T+\tau_{k}}^{(i-\mathbbm{1}_{m>k})T+\tau_{m}} \!\!\!\!h^\prime_m(t)\,dt, \nonumber\\
\label{eq:A2}
\hspace{-0.65cm}&& c^\prime_{m,k}[i] = \frac{1}{T} \int_{(i-\mathbbm{1}_{m>k})T+\tau_{m}}^{iT+\tau_{k}} h'_m(t) e^{j\varepsilon^\prime_m \zeta}\,d\zeta = \frac{1}{j\varepsilon^\prime_m T}\times \\
\hspace{-0.65cm}&& \left(e^{j\varepsilon^\prime_m[iT+\tau_{k}]}-e^{j\varepsilon^\prime_m(i-\mathbbm{1}_{m>k})T+\tau_{m}}\right)
\int_{(i-\mathbbm{1}_{m>k})T+\tau_{m}}^{iT+\tau_{k}} \!\!\!\!h'_m(t)\,dt.\nonumber
\end{eqnarray}
\normalsize

On the other hand, for our WMFS scheme, the $y_k[i]$ samples in \eqref{eq:samples} can be written to
\begin{eqnarray}
y_k[i] =  \sum_{m=1}^M g_{m,k}[i] s_m[i-\mathbbm{1}_{m>k}] + \tilde{z}_k[i],
\end{eqnarray}
where
\begin{eqnarray*}
g_{m,k}[i]
\hspace{-0.2cm}&=&\hspace{-0.2cm} h'_m \frac{1}{d_k} \int_{(i-1)T+\tau_{k}}^{(i-1)T+\tau_{k+1}} e^{j\varepsilon'_m \zeta}d\zeta \\
\hspace{-0.2cm}&=&\hspace{-0.2cm}  \frac{h'_m}{d_k} \frac{e^{j\varepsilon'_m d_k}-1}{j\varepsilon'_m} e^{j\varepsilon'_m[(i-1)T+\tau_k]}.
\end{eqnarray*}

As a result, if the channel is fast fading and there is residual CFO in the received signal, we can simply modify the coefficient matrix $\bm{D}$ as
\begin{eqnarray}\label{eq:A3}
&&\hspace{-0.7cm} \bm{D}=\\
&&\hspace{-0.7cm} \begin{bmatrix}
\begin{smallmatrix}
g_{1,1}[1] &            &       &             &             &               &    &               &    &     \\
g_{1,2}[1] & g_{2,1}[1] &       &             &             &               &    &               &    &     \\
...        & g_{2,2}[1] & ...   &             &             &               &    &               &     &   \\
g_{1,M}[1] & ...        & ...   &  g_{M,1}[1] &             &               &    &               &    &    \\
           & g_{2,M}[1] & ...   &  g_{M,2}[1] & g_{1,1}[2]  &               &    &               &    &    \\
           &            & ...   &  ...        & g_{1,2}[2]  & g_{2,1}[2]    &    &               &     &   \\
           &            &       &  g_{M,M}[1] & ...         & g_{2,2}[2]    & ...  &             &    &    \\
           &            &       &             & g_{1,M}[2]  & ...           & ...  & g_{M,1}[2]  &    &    \\
           &            &       &             &             & g_{2,M}[2]    & ...  & g_{M,2}[2]  & ...&       \\
           &            &       &             &             &               & ...  & ...         & ... &   \\
           &            &       &             &             &               &      & g_{M,M}[2]  & ... &     \\
           &            &       &             &             &               &      &             & ... &
\end{smallmatrix}
\end{bmatrix}. \nonumber
\end{eqnarray}

\section{Additional Simulation Results}\label{sec:AppB}
\subsection{Channels with mismatched amplitudes}
In Section~\ref{sec:V}, we only consider phase-misaligned channels, but it is easy to generalize the simulations to the misaligned-amplitude case. Below, we repeat the simulation in Fig. 8, assuming mismatched amplitudes of the residual channel gains.

As in Fig. \ref{fig:sim3}, we fix the maximum time offset to $\tau_M=0.5$ in the simulation. The residual channel gain of the $m$-th device is $h_m=|h_m|e^{j\phi_m}$. In particular, the phase term is uniformly distributed, i.e., $\phi_m\sim U(0,\phi)$, where $\phi$ is fixed to $\pi/2$. Unlike Fig. 8, where $|h_m|=1$, we assume $|h_m|$ follows the Rayleigh distribution, i.e., $|h_m|\sim\text{Rayleigh}(\sigma)$, where $\sigma$ is set to $0.25$, $0.5$, and $1$, respectively.

\begin{figure}[t]
  \centering
  \includegraphics[width=0.8\columnwidth]{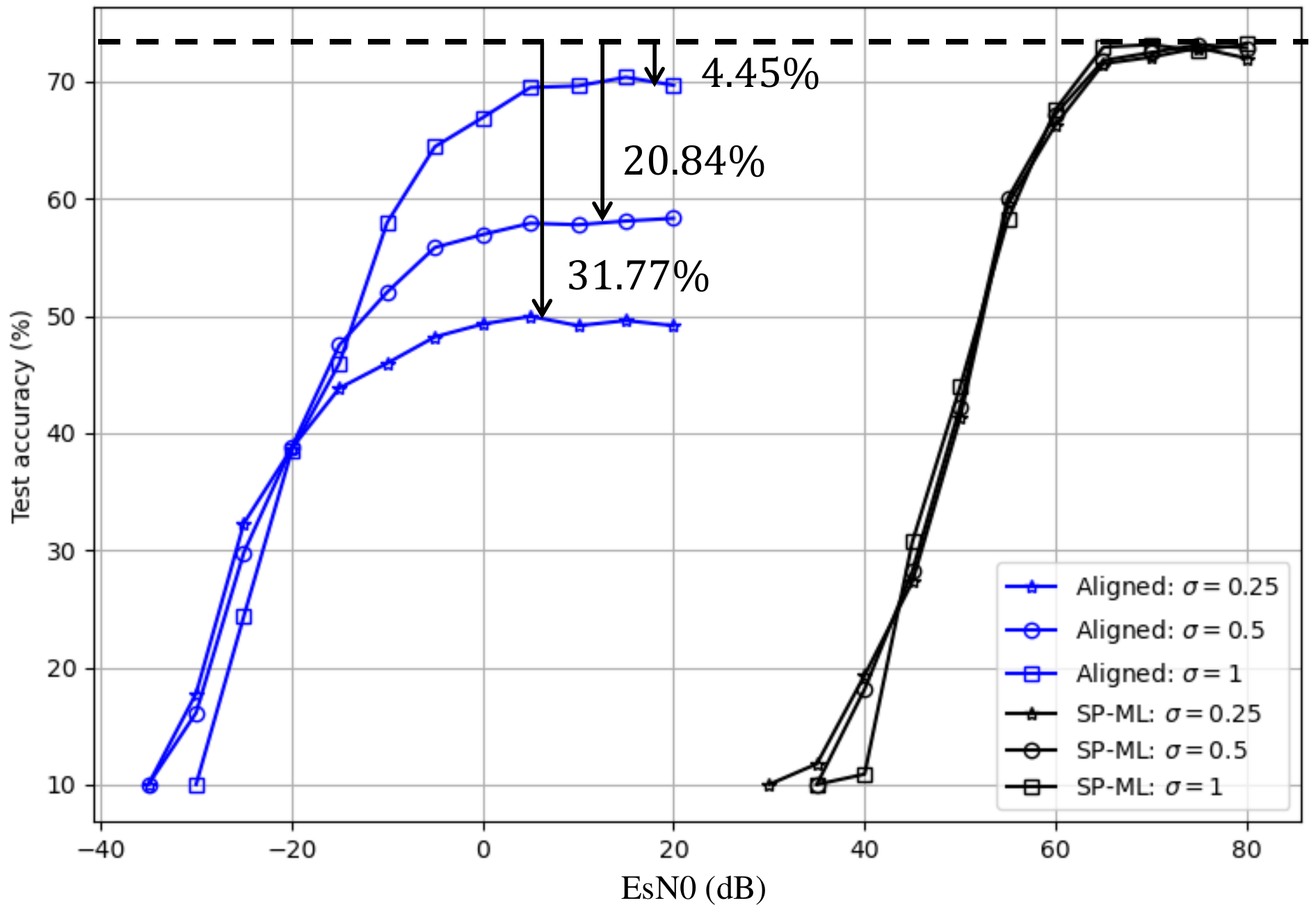}\\
  \caption{Test accuracies of the aligned-sample and ML estimators under different degrees of channel misalignments. The maximum time offset is $\tau_M=0.5$, the maximum phase misalignment $\phi=\pi/2$, and the amplitude of the residual channel is $|h_m|\sim\text{Rayleigh}(\sigma)$, where $\sigma$ is set to $0.25$, $0.5$, and $1$, respectively.}
\label{fig:R1}
\end{figure}

Fig. \ref{fig:R1} presents the test accuracies of the learned model in the presence of mismatched channel amplitudes, where both the aligned-sample and SP-ML estimators are used. In general, the observations from Fig. \ref{fig:R1} match those from Fig. \ref{fig:sim3}. Specifically,
\begin{enumerate}
\item The aligned-sample estimator works well in the low-EsN0 regime while the ML estimator works well in the high-EsN0 regime.
\item Due to the channel-gain misalignment in both amplitude and phase, the aligned-sample estimator suffers from a test-accuracy loss. On the other hand, the ML estimator does not suffer from test-accuracy loss -- the test accuracy after convergence is the same as the aligned case.
\end{enumerate}

\subsection{Effect of the number of active devices}
In Section~\ref{sec:V}, we consider a FEEL system with $M = 40$ edge devices. $50,000$ training examples are assigned to the $40$ devices in a non-i.i.d. manner. In each FEEL iteration, a random subset of $K$ devices are active and $K$ is set to $4$ in the simulations. In the following, we further investigate the impact of $K$ on the learning performance.

\begin{figure*}[t]
  \centering
  \includegraphics[width=1.25\columnwidth]{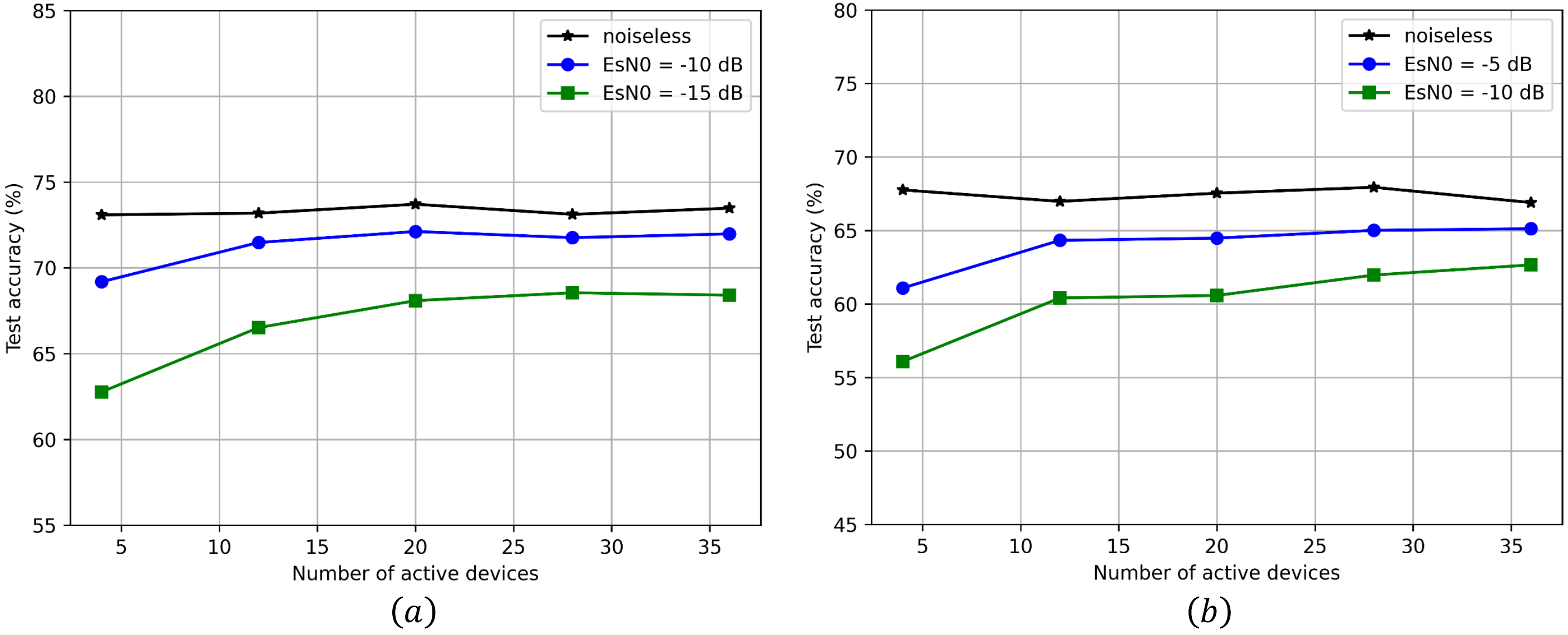}\\
  \caption{Test accuracy versus the number of active devices in FEEL, where $M = 40$. (a) Aligned OAC, where there is neither time nor channel-gain misalignment, only AWGN is introduced in the federated averaging process. (b) Misaligned OAC, where the maximum time offset $\tau_M=0.5$ and the maximum phase misalignment $\phi=\pi/2$. We use the aligned-sample estimator at the PS to estimate the aggregated model.}
\label{fig:R2}
\end{figure*}

Fig. \ref{fig:R2} presents the learning performance of FEEL versus the number of active devices considering both aligned and misaligned OAC. Let us first focus on the aligned OAC, in which case only AWGN is introduced in the model aggregation process and there are no misalignments. The test-accuracy performance is shown in Fig. \ref{fig:R2}(a).
\begin{enumerate}
\item A first observation is that the learning performance is irrelevant to the number of active devices in each iteration when the noise power is zero (i.e., the noiseless case, where FEEL reduces to federated learning (FL) with ideal model aggregation). We conjecture that this is due to the non-i.i.d. data across devices. Similar results are observed in the convergence proof of FL \cite{li2019convergence}, where the authors showed that the number of active devices has very limited influence on the convergence of FL when the data are non-i.i.d..
\item On the other hand, when there is noise, the learning performance benefits from more participating devices in each FEEL iteration. In general, more participating devices yield a smaller variance of the aggregated gradients, thereby stabilizing the learning in a noisy environment.
\end{enumerate}

\begin{figure*}[t]
  \centering
  \includegraphics[width=1.25\columnwidth]{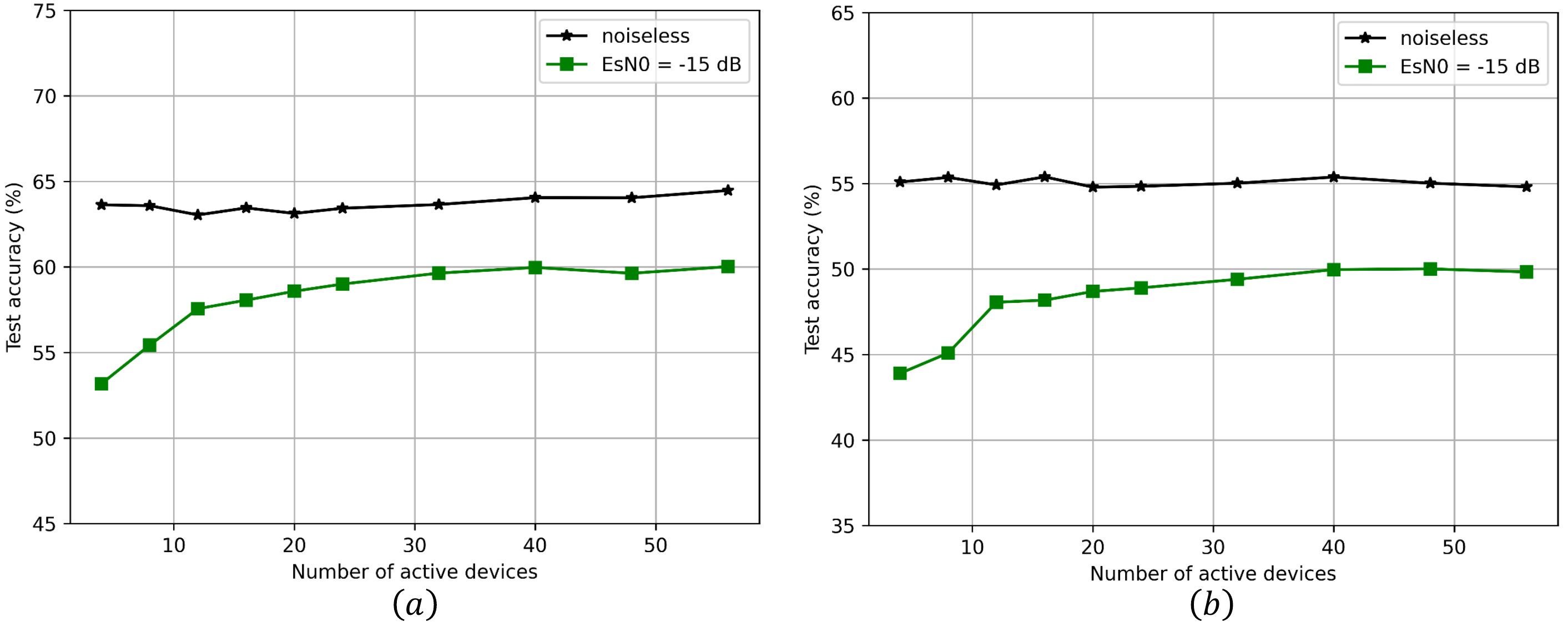}\\
  \caption{Test accuracy versus the number of active devices in FEEL, where $M = 80$. (a) Aligned OAC, where there is only AWGN but no time or channel-gain misalignment. (b) Misaligned OAC, where the maximum time offset $\tau_M=0.5$, the maximum phase misalignment $\phi=\pi/2$, and the aligned-sample estimator is used to estimate the aggregated model.}
\label{fig:R3}
\end{figure*}

When there are misalignments in the overlapped signals, the same results can be observed, as shown in Fig. \ref{fig:R2}(b). Specifically, when there are only misalignments but no noise, the learning performance is relatively stable with the increase in the number of active devices. When there are both misalignments and noise, however, more participating devices are beneficial to the learning performance.

We further performed larger-scale simulations with more than $40$ devices, as required by the reviewer, and present the results in Fig. \ref{fig:R3} below. In these simulations, we assume there are $M = 80$ edge devices in the FEEL system and the number of active devices in each iteration, $K$, ranges from $4$ to $56$. 
As can be seen, the results are in line with those in Fig. \ref{fig:R2}.

\bibliographystyle{IEEEtran}
\bibliography{References}

\end{document}